\documentclass[12pt]{article}
\pdfoutput=1
\usepackage[T1]{fontenc} 
\usepackage[style=ieee,citestyle=numeric,doi=false]{biblatex}
\renewbibmacro*{bbx:savehash}{}

\usepackage[english]{babel}
\usepackage{graphicx}
\usepackage{amsmath}
\usepackage{amssymb}
\usepackage{multirow}
\usepackage{url}
\usepackage{tikz}
\usepackage{braket}
\usetikzlibrary{decorations.markings}
\usepackage{subcaption}

\newcommand\tphi{\tilde{\phi}}
\usepackage{hyperref}
\usepackage{calrsfs}
\DeclareMathAlphabet{\pazocal}{OMS}{zplm}{m}{n}

\allowdisplaybreaks
\def\CR{\nonumber \\}
\def\eq#1{(\ref{#1})}
\def\[#1\]{\begin{align}#1\end{align}}

\begin{document}

\begin{titlepage}

\title{
\hfill\parbox{4cm}{ \normalsize YITP-17-107}\\ 
\vspace{1cm} 
Emergent symmetries in the canonical tensor model
}
\author{
Dennis Obster$^{1,2}$\footnote{dobster@science.ru.nl}, 
Naoki Sasakura$^2$\footnote{sasakura@yukawa.kyoto-u.ac.jp}
\\
$^1${\small{\it Institute for Mathematics, Astrophysics and Particle Physics, Radboud University,}}
\\ {\small{\it Heyendaalseweg 135, 6525 AJ Nijmegen,The Netherlands}}
\\
$^2${\small{\it Yukawa Institute for Theoretical Physics, Kyoto University,}}
\\ {\small{\it  Kitashirakawa, Sakyo-ku, Kyoto 606-8502, Japan}}
}

\date{\today}

\maketitle

\begin{abstract}
The canonical tensor model (CTM) is a tensor model proposing a 
classically and quantum mechanically consistent model of gravity, formulated as a first-class
constraint system with structural similarities to the ADM formalism of general relativity.
A recent study on the formal continuum limit of the classical CTM has shown that it produces a general relativistic system.  
This formal continuum limit assumes the emergence of a continuous space, but ultimately
continuous spaces should be obtained as preferred configurations of the quantum CTM. 
In this paper we study the symmetry properties of a wave function
which exactly solves the quantum constraints of the CTM for general $N$. 
We have found that it has strong peaks at configurations invariant under some Lie-groups,
as predicted by a mechanism described in our previous paper.
A surprising result was the preference of configurations invariant not only under Lie-groups 
with positive signatures, but also with spacetime-like signatures, i.e.,\,$SO(1,n)$.
Such symmetries could characterize the global structures of spacetimes,
and our results are encouraging towards showing spacetime emergence in the CTM.
To verify the asymptotic convergence of the wave function we have also analyzed the asymptotic behaviour, which for the most part seems to be well under control. 
\end{abstract}

\end{titlepage}

\section{Introduction}
 The current standard model of particle physics describes three of the fundamental forces with great precision. Perturbative methods in quantum field theory are used to calculate scattering amplitudes of processes. The notable absentee in this description of fundamental physics is gravity. The absence of gravity is due to the perturbative non-renormalizability of Einstein's general relativity~\cite{GOROFF1986709}, making the perturbative theory lose its predictive power. Ever since there have been attempts to develop well-defined non-perturbative theories which lead to general relativity in some continuum limit.
 
 One way to treat quantum gravity non-perturbatively is by introducing a discretization of spacetime by means of simplices at the Planck scale. One of the ways to do this is by the use of tensor 
 models~\cite{Sasakura:1990fs,Ambjorn:1990ge,Godfrey:1990dt}, which can be seen as a generalization of matrix models. These original models are known to have some 
difficulties\footnote{A serious problem of the tensor models with 
 symmetric tensors like the original models is that it is unknown whether there exist $1/N$ expansions which would enable
systematic analysis. Recently, introducing a traceless condition~\cite{Klebanov:2017nlk} 
or a pair of symmetric tensors~\cite{Gurau:2017qya} has been proposed as possible resolutions.}, 
some of which have been resolved by the advent of the colored tensor models~\cite{Gurau:2009tw}.
However, there still remain problems due to the emergence of branched polymers instead of 
space-like simplicial complexes resembling our universe\footnote{Recently, tensor models 
are attracting much attention as SYK-like models without disorder~\cite{Witten:2016iux,Klebanov:2016xxf}. In this context, tensor models may be indirectly related to quantum gravity through holography,
in which the dominance of branched polymer-like graphs, the so-called Melonic diagrams, is important in the exact solvability of the model in the large-$N$ limit.}~\cite{Gurau:2013cbh,Bonzom:2011zz}. 
The problem here may lie in the fact that tensor models usually by construction generate Euclidean signature spaces, without paying special attention to time. One model which treats time differently is causal dynamical triangulation and it is able to produce macroscopic spaces using a notion of causality to restrict generated spaces to be compatible with a (3+1)-dimensional Lorentzian decomposition~\cite{Ambjorn:2004qm,Ambjorn:2012jv}. On the other hand, the Euclidean counterpart (dynamical triangulation) has proven to be more difficult~\cite{Coumbe:2014nea, Laiho:2016nlp}.
 
 The issues of the original tensor models and the suggestion of the importance of the treatment of time in quantum gravity led to a model called the canonical tensor model (CTM), which has been introduced by one of the authors of this paper~\cite{Sasakura:2011sq}. The model is defined in the Hamiltonian (also called the canonical) formalism, which naturally treats time separately. Like the Hamiltonian formulation of general relativity (the ADM~formalism~\cite{Arnowitt:1962hi}), the Hamiltonian consists of a linear combination of first class constraints. This makes sure that, even though time is singled out, general covariance is not broken. The fundamental dynamical variables of the model are a conjugate pair of real symmetric rank-3 tensors. The CTM has been shown to have a strong connection to general relativity: It agrees with a mini-superspace approximation for $N=1$~\cite{Sasakura:2014gia},\footnote{Here, $N$ denotes the range of the indices of the tensors, namely $1,2,\cdots,N$.} while in a formal continuum limit, where $N\rightarrow\infty$, the dynamical structure agrees with that of general relativity~\cite{Sasakura:2015pxa,Chen:2016ate}. Due to this connection with general relativity and the fact that the model can be quantized easily~\cite{Sasakura:2013wza}, one can hope for this model to be a consistent model for quantum gravity. 
  
 Since the main goal of the CTM is to describe quantum gravity, it is important to study the  quantum mechanical dynamics of the model, e.g., the physical states (wave functions)~\cite{Sasakura:2013wza, Narain:2014cya}. One important question is what the properties of the preferred configurations of the wave functions are. Especially symmetries are interesting to analyze since they might give a hint as for what kind of spaces can emerge from the model. In this paper we analyze these preferred configurations of a wave function  of the model which is valid for general $N$, 
particularly paying attention to the mechanism described before in~\cite{Obster:2017pdq}, where configurations which are themselves invariant under a subgroup of the full symmetry group of a system get amplified.\footnote{The main point of this mechanism is that the physical quantity describing a state in the system is invariant under a group $G$, whereas certain configurations can themselves be invariant under a subgroup $H\subset G$. It was found that these configurations will be greatly preferred over non-symmetric configurations.} We find that this mechanism seems to work well for this wave function of the CTM, observing clear preferences of symmetric configurations. 
Rather surprisingly, we find not only Lie-groups with space-like (positive definite) signatures, 
but also the ones with spacetime-like (indefinite) signatures, particularly $SO(1,n)$,  
as the symmetries associated with the preferred configurations.
This suggests that there is even a hidden time direction present in the emergent symmetry, 
possibly signaling the emergence of deSitter-like spacetimes in the CTM.\footnote{Some discretion with this statement is in order here. 
Though the appearance of such spacetime symmetries is encouraging,
a correct geometrical interpretation is still left for later study.}
 
This paper is organized as follows. In Section~\ref{sec:review}, we review the formalism of the CTM and present
the wave function we analyze. The wave function is expressed as a holomorphic integration over $N$ variables, which is a sort of a multi-variable generalization of the Airy function.
In Section~\ref{sec:mechanism}, we review the mechanism of the amplification of 
the wave function at symmetric configurations, 
which was described in our previous paper \cite{Obster:2017pdq}. In Section~\ref{sec:general},
we explain the method by which we numerically evaluate the wave function at generic configurations.
Since the integrand is oscillatory (oscillating infinitely fast at infinity) with a constant modulus
we introduce a regularization procedure, which in physics is often called the $\epsilon$-prescription, to properly handle the 
conditionally convergent integral.
Then, we take the vanishing limit of the regularization 
by considering a deformation of the integration contour.
We introduce a numerical method which takes care of the deformation.
In Section~\ref{sec:simplified}, we consider a subspace of the configurations, 
in which one can analytically carry out all the integrations except for one. This simplified model is
useful for studying the amplification mechanism, especially for large-$N$ cases, 
because only one numerical integration is necessary for any $N$. 
We observe strong amplification of the wave function at symmetric configurations. 
We also study some large-$N$ behavior of the wave function.    
In Section~\ref{sec:timelike} we show how the space-like symmetries which are highlighted are promoted to spacetime-like symmetries in this wave function.
In Section~\ref{sec:asymptotic}, we study the asymptotic behavior of the wave function at the infinity
of the configuration space, numerically and analytically.  We find a rich variety of behaviors, which should be 
studied more thoroughly in the future. 
The final section is devoted to a summary and future problems.

\section{Review of the canonical tensor model}
\label{sec:review}
The canonical tensor model (CTM) is a model for gravity in the canonical (Hamiltonian) framework, which seems to be a natural starting point to construct a model which treats time differently. There have been several attempts to do this by starting from the ADM formalism~\cite{Arnowitt:1962hi}
of general
relativity, where it is described as a first-class constrained system with the fundamental fields being the spatial metric $h_{ij}$ and its conjugate momentum $\pi^{ij}$. The (reduced) Hamiltonian density $\mathcal{H}$ is given by a linear combination of the so-called ``Hamiltonian constraint'' $\pazocal{H}$ and the  ``(spatial) diffeomorphism constraint'' $\pazocal{H}_i$,\footnote{The diffeomorphism constraint is also often called the momentum constraint.} where the lapse function $N$ and the shift vector $N^i$ act like the corresponding Lagrange multipliers:\footnote{For a geometric overview of these quantities, see for instance~\cite{wald1984general}.}
 \begin{equation}
     \mathcal{H} = N \pazocal{H}+N^i \pazocal{H}_i.\label{eq:review:H_ADM}
 \end{equation}
 The constraints span the \emph{hypersurface deformation algebra},
 \begin{align}
     \begin{split}
          \{ H(f), H(f') \} &= \vec{H}(\vec{F}), \\
         \{ \vec{H}(\vec{f}), H(f) \} &= H(\pazocal{L}_{\vec{f}} f),\\
       \{ \vec{H}(\vec{f}), \vec{H}(\vec{f'}) \} &= \vec{H}(\pazocal{L}_{\vec{f}} \vec{f'}),
     \end{split}\label{eq:review:ADM_algebra}
 \end{align}
 where $\vec{H}(\vec{f}) = \int d^3x f^i \pazocal{H}_i$, $H(f) = \int d^3x f \pazocal{H}$, $F^i = h^{ij} (f\partial_j f' - f' \partial_j f)$ and $\pazocal{L}_{\vec{f}}$ is the Lie derivative with respect to $\vec{f}$.
 
 The most straightforward way to attempt to construct a canonical quantum gravity theory is to quantize the fundamental fields
 by mapping $h_{ij}\rightarrow \hat{h}_{ij}$ and $\pi^{ij}\rightarrow\hat{\pi}^{ij}.$\footnote{This does not appear to be the best method in canonical quantum gravity and one is better off using Ashtekar variables~\cite{Ashtekar:1986yd} which led to the loop representations in quantum gravity (Loop Quantum Gravity).}
Since $\pazocal{H}$ and $\pazocal{H}_i$ are classical constraints, one can implement them on the quantum level by demanding
 \begin{align}
     \hat{\pazocal{H}} \ket{\Psi} &= 0,\label{eq:review:wheelerdewitt}\\
     \hat{\pazocal{H}}_i \ket{\Psi} &= 0.\label{eq:review:momentumconstraint}
 \end{align}
 Here,~\eqref{eq:review:wheelerdewitt} is called the Wheeler-deWitt equation. This functional differential equation is in general not well-defined, although some attempts have been made to make sense of this. There are numerous difficulties in this approach with
 varieties of seriousness.
 
 To circumvent these issues in the canonical formalism, one may try to describe a space in a discrete way by a set of ``points''. In the CTM we choose to implement this already at the classical level, by describing the model as a tensor model. The first non-trivial case to try to construct a Hamiltonian with the similar properties as \eqref{eq:review:H_ADM} would be a real symmetric
 rank-3 tensor model, using a conjugate pair of real symmetric rank-3 tensors, $Q_{abc}$ and $P_{abc}$, as the fundamental variables with the canonical Poisson algebraic relations,
 \begin{align}
 \begin{split}
     \{ Q_{abc}, P_{def} \} &= \sum_{\sigma} \delta_{a\sigma_d}\delta_{b\sigma_e}\delta_{c\sigma_f} ,\\
     \{Q_{abc}, Q_{def} \}  &= \{P_{abc}, P_{def} \} = 0,
\end{split}
\label{eq:fundamental}
 \end{align}
 where $\sigma$ are the permutations of $d,e$ and $f$.
 The labels of the tensors range from $1$ to $N$ and 
 label the ``points'' in the space we are interested in\footnote{We call them ``points'', since the formal continuum limit suggests that the labels are mapped to continuous coordinates. The exact implementation of this geometric picture at the discrete level is still not fully understood, as the points should be connected in some way, e.g., by simplices.}. The $N\rightarrow\infty$ limit is supposed to 
 correspond to a continuous space where the model should coincide with general 
 relativity.\footnote{It is worth stressing that unlike the usual Euclidean-type tensor models, 
 the spacetime dimension emerging from the CTM is not directly related to
 the rank of the tensors.
 This can for instance be seen in the actual correspondence in 
 a formal continuum limit~\cite{Sasakura:2015pxa,Chen:2016ate}.} 
 Similar to the spatial diffeomorphism invariance in general relativity, we introduce a kinematical $O(N)$ symmetry of the system such that the system is invariant under ``relabeling'' of the points:
 \[
   \begin{split}
      Q_{abc} &\rightarrow L_{aa'} L_{bb'} L_{cc'} Q_{a'b'c'},\\
      P_{abc} &\rightarrow L_{aa'} L_{bb'} L_{cc'} P_{a'b'c'}.
   \end{split}\label{eq:review:ON_invariance}
 \] 
 Here the $L_{ab}$ are $O(N)$ matrices. The CTM is the minimal of its kind, meaning that we consider a model with just two constraints:
 \begin{equation}
     H = n_a\pazocal{H}_a + n_{ab} \pazocal{J}_{ab},
 \end{equation}
 where $\pazocal{H}_a$ corresponds to the Hamiltonian constraint and $\pazocal{J}_{ab}$ corresponds to the spatial diffeomorphism constraint in \eqref{eq:review:H_ADM}. 
For convenience, and to maintain the analogy to the ADM formalism, we will abuse this terminology to refer to the CTM constraints from now on. 
The spatial diffeomorphism constraint of ADM generates diffeomorphisms within a certain timeslice~\cite{thiemann2007modern}, 
hence it is natural to take for $\pazocal{J}_{ab}$ the generators of $SO(N)$ transformations we imposed in \eqref{eq:review:ON_invariance}:
 \begin{equation}
     \pazocal{J}_{ab} = \frac{1}{4}(Q_{acd}P_{bcd} - Q_{bcd}P_{acd}),\label{eq:review:J_constraint}
 \end{equation}
 which is anti-symmetric, $\pazocal{J}_{ab}=-\pazocal{J}_{ba}$.
 To specify the classical model, one now has to introduce the Hamiltonian constraint. 
 In analogy with the ADM formalism of
 general relativity, the algebra spanned by the constraints should close. Furthermore one can deduce that the terms should be connected, corresponding to the absence of non-local behaviour in the ADM algebra \eqref{eq:review:ADM_algebra}. This means that for instance a term like $Q_{abc}Q_{bcd}Q_{dee}$ is allowed but a term like $Q_{abb}Q_{cde}Q_{cde}$ is not. By considering only terms which are up to the third order in $Q$ and $P$ and even in $P$,\footnote{The limitation up to the third order has been put in by hand in order to consider the simplest case, 
and we do not know whether there exist other consistent Hamiltonian constraints with 
higher order terms.  
Whether it is fair to only consider even terms in $P$ is still an open question; however, it can be somewhat physically motivated as a ``time reversal symmetry'' condition where $P\rightarrow-P$ and $Q\rightarrow Q$.} one can prove that there is a unique model described by the following Hamiltonian constraint~\cite{Sasakura:2012fb}:
 \begin{equation}
     \pazocal{H}_a = \frac{1}{2} (P_{abc} P_{bde} Q_{cde} - \lambda\, Q_{abb}),\label{eq:review:H_constraint}
 \end{equation}
 where $\lambda$ is a real constant. Without loss of generality, the constant can be normalized 
 as $\lambda=0,\pm1$ by a rescaling, $Q\rightarrow c\, Q,\ P\rightarrow P/c$,
 which keeps \eq{eq:fundamental}.
 The constraints of \eqref{eq:review:J_constraint} and \eqref{eq:review:H_constraint} span the following algebra, which corresponds to the ADM algebra in a formal continuum limit with
 $N\rightarrow\infty$ \cite{Sasakura:2015pxa}:
 \begin{align}
     \begin{split}
       \{ \pazocal{H}(\xi^1), \pazocal{H}(\xi^2) \} &= \pazocal{J}\left([\tilde{\xi}^1, \tilde{\xi}^2] + 2 \lambda\, \xi^1 \wedge \xi^2\right),\\
       \{\pazocal{J}(\eta) , \pazocal{H}(\xi)\} &= \pazocal{H}(\eta \xi),\\
       \{ \pazocal{J}(\eta^1), \pazocal{J}(\eta^2) \} &= \pazocal{J}\left([\eta^1, \eta^2]\right).
     \end{split}
 \end{align}
 Here $\pazocal{H}(\xi) = \pazocal{H}_a \xi_a$, $\pazocal{J}(\eta) = \pazocal{J}_{ab} \eta_{ab}$, $\tilde{\xi}_{ab} = P_{abc}\xi_c$, $(\xi^1\wedge\xi^2)_{ab} = \xi_a\xi_b - \xi_b\xi_a$ and $[.,.]$ denotes the matrix commutator.
 
 The quantization of the CTM can be done consistently by canonical quantization~\cite{Sasakura:2013wza}. Let us map the canonical variables to quantum mechanical operators and the canonical Poisson brackets to quantum mechanical commutators
 \begin{align}
  \begin{split}
   & Q_{abc} \rightarrow \hat{Q}_{abc}, P_{abc} \rightarrow \hat{P}_{abc},\\
    & \{ Q_{abc}, P_{def} \} \rightarrow -i [ \hat{Q}_{abc}, \hat{P}_{def} ].
  \end{split}
 \end{align}
 The constraints are now given by the operators
  \begin{align}
      \hat{\pazocal{H}}_a &= \frac{1}{2} (\hat{P}_{abc} \hat{P}_{bde} \hat{Q}_{cde} - \lambda \hat{Q}_{abb} + i\lambda_H \hat{P}_{abb}),
      \label{eq:quantumHamiltonian}\\
      \hat{\pazocal{J}}_{ab} &= \frac{1}{4}(\hat{Q}_{acd} \hat{P}_{bcd} - \hat{Q}_{bcd} \hat{P}_{acd}).
  \end{align}
  The constant $\lambda_H$ depends on the ordering of the operators in the
  first term of the Hamiltonian constraint. However,  if one requires the Hamiltonian constraint
  to be self-adjoint, this constant is fixed to be 
  \[
  \lambda_H = \frac{1}{2}(N+2)(N+3).
  \label{eq:valoflambdah}
  \]
  Conveniently, the quantized constraint algebra contains no anomalies:
  the algebra remains of the same form, as can be checked by explicit computations. 
  
  Just like the usual constraints in canonical quantum gravity, \eqref{eq:review:wheelerdewitt} and \eqref{eq:review:momentumconstraint}, we have to impose that the physical states of the theory vanish under the constraints
  \begin{align}
      \hat{\pazocal{H}}_a \ket{\Psi} &= 0,\label{eq:review:CTMwheelerdewitt}\\
      \hat{\pazocal{J}}_{ab} \ket{\Psi} &= 0.\label{eq:review:CTMmomentum}
  \end{align}
  By choosing a representation, these constraints can be expressed as a set of partial differential equations. This means that the problem is in principle well-defined, though the solutions can in general have very complicated forms. Several exact solutions have been found before~\cite{Sasakura:2013wza, Narain:2014cya}, and the main interest of this paper is a wave function in the $P$ representation given by
  \begin{equation}
      \Psi(P) = \int_{{\mathbb R}^{N+1}} d\phi d\tphi\  e^{i (P \phi^3 + \phi^2 \tphi- \frac{4}{27 \lambda} \tphi^3)},\label{eq:review:Psi}
  \end{equation}
  which gives the wave function of a physical state by 
  \[ 
  \Psi_{phys}(P)=\Psi(P)^\frac{\lambda_H}{2}.
\label{eq:relpsiQP}
\]
Here $\phi=(\phi_1,\phi_2,\ldots,\phi_N) \in \mathbb{R}^N$, $\tphi\in\mathbb{R}$ and we use the short-hand notations,
  \begin{align}
     \begin{split}
         P\phi^3 &= P_{abc} \phi_a \phi_b \phi_c,\\
         \phi^2 &= \phi_a \phi_a, \\
         d\phi &= \Pi_{a} d\phi_a.
     \end{split} 
  \end{align}
  Though there are several exact solutions known, \eq{eq:review:Psi} has the nice property that it is valid for any $N$.
  The derivation of this wave function is given in Appendix~\ref{app:wavefunction}.
  We will mainly consider the $\lambda>0$ case, 
  as it appears to be the physically most sensible case.
  Taking $\lambda\rightarrow 0$ (after rescaling $\tilde \phi\rightarrow |\lambda|^{1/3}\tilde \phi$
  in \eq{eq:review:Psi})
  will lead to the problems mentioned below.
 The problem of $\lambda<0$ will be shown in Section~\ref{sec:general}.
  The wave function with $\lambda>0$ will in fact be shown to have rich structures, 
  which can be well understood in terms of 
  the highlighting mechanism of symmetries reviewed in Section~\ref{sec:mechanism}. 
  As discussed in Section~\ref{sec:asymptotic} and Appendix~\ref{app:normalizable}, the wave function seems
  asymptotically decaying and normalizable in most of the directions of the $|P|\rightarrow\infty$ limit, 
  so it seems reasonable to concentrate our attention mainly to 
  the structure of the peaks in the finite region of $P$.
  
 As for the $\lambda=0$ case,  another wave function is known which is
 generally valid for any $N$ in the $Q$-representation \cite{Narain:2014cya}\footnote{This wave function is a solution to the second order partial differential equations 
 derived from the Hamiltonian constraints in the $Q$-representation. 
 Therefore, the solution is more non-trivial than \eq{eq:review:Psi} in the $P$ representation, which is a solution
 to the first order ones. Unfortunately, we do not presently have any generalization of this solution 
 to $\lambda\neq0$.} and is given by
  \begin{equation}
      \Psi(Q) = \int_{\mathbb{R}^{N}} d\phi\  (\phi^2)^{\alpha} e^{i Q \phi^3},
      \label{eq:exactwithnolambda}
  \end{equation}
  where $\alpha=(N+3)(N-2)/8$. 
  The idea to look at $\lambda=0$ seems physically justified, since the $\lambda$-term 
  in \eq{eq:quantumHamiltonian} classically leads to non-local behaviour in the formal continuum limit \cite{Sasakura:2015pxa,Chen:2016ate}.
  However, as shown in Appendix~\ref{appendix:sec:wavewithoutl},
  the wave functions, \eq{eq:review:Psi} with $\lambda=0$ and \eq{eq:exactwithnolambda},
 have some singular behaviors which would make sensible interpretations difficult. 
  In short, for $\lambda=0$, 
  due to the homogeneous nature of the wave functions under the rescaling of $P$ (or $Q$),
  the system suffers from an instability of collapsing down to 
  vanishing (or divergent) configurations, namely $P=0$ (or $Q=\infty$). 
  On the other hand, as we will see, the wave function \eq{eq:review:Psi} for $\lambda>0$ has 
  some interesting behavior for finite $P$, which is potentially of physical importance.  
  Thus $\lambda>0$ seems to be the only physically sensible choice\footnote{$\lambda$ 
  corresponds to the cosmological constant in the correspondence between 
  the CTM with $N=1$ and the minisuperspace 
  treatment of general relativity~\cite{Sasakura:2014gia}. 
  Therefore, the necessity of $\lambda>0$ is curiously matching
  the present astrophysical observation of a positive cosmological constant.}
  in the quantum CTM.
 This also suggests an important future subject of study 
 that the classical dynamics analyzed for 
 $\lambda=0$ in~\cite{Sasakura:2013wza, Narain:2014cya} should get modified 
 from this quantum requirement.
  
  Lastly, we would like to comment on the Lorentzian form we have particularly
taken in \eq{eq:review:Psi},
 namely the exponent of the integrand has an overall factor $i$
  with a real action\footnote{For convenience we use the terminology `action', 
  inspired by the path integral formulation of quantum field theories.
  The action  in \eq{eq:review:Psi} is $S=P \phi^3 + \phi^2 \tphi- \frac{4}{27 \lambda} \tphi^3$.} 
  and the integration region is a real hyper-plane. 
  As reviewed in Appendix~\ref{app:wavefunction},
only the validity of partial integrations is the essential ingredient of the proof for the wave 
function to be the solution. 
  Therefore, as far as the integration is convergent, one can freely take the overall factor and 
  the integration region:  There is no particular reason to take the Lorentzian form from 
  the requirement of physical states.
 On the other hand, in our
treatment of paying special attention to the time direction,
the Lorentzian form would be the most natural choice, and 
moreover it has 
 the following two advantages.
One is that the wave function is well-defined (at least for non-special values of $P$) as is.
As will be discussed in later sections, the wave function is generally well-defined as a conditionally
convergent integral for generic $P$. This will pose more difficult problems in the Euclidean form,
as the integral will usually suffer from divergences caused by the cubic terms, unless the 
integration region is altered to some non-trivial complex one.
Another important reason is that the highlighting mechanism of symmetries
explained in Section~\ref{sec:mechanism}
requires the coherence/de-coherence of the integrand to occur, and 
the Lorentzian form would be the most efficient one for the mechanism to be evident. 
Therefore, though it might be theoretically possible to make some other choices, we will 
exclusively consider the Lorentzian form throughout this paper.

\section{Highlighting mechanism of symmetries}\label{sec:mechanism}
 In~\cite{Obster:2017pdq} the authors of this paper introduced a mechanism which can explain the preference of symmetric configurations in models similar to the CTM. This section serves as a short review of this mechanism.
 
 In physics, one is often interested in quantities of the form
 \begin{equation}
      \Psi(Q) = \int_{\pazocal{C}} d\phi\  e^{i S(\phi, Q)}.\label{eq:mech:genPsi}
 \end{equation}
 Here $\Psi$ is the physical quantity of interest, defined on a configuration space of which $Q$ is an element. $\phi$ denotes some internal integration variables in the space $\pazocal{C}$ and $S(\phi, Q)$ is a functional of $\phi$ and $Q$. Note that the wave function \eqref{eq:review:Psi} of our interest also has this form. 
 
 Let us introduce a group symmetry in \eq{eq:mech:genPsi}.
 Considering $Q$ and $\phi$, which are labelled by some discrete set of labels,
 one can look at the action of some group $G$ on \eqref{eq:mech:genPsi}, where $Q$ and $\phi$ transform under some representations,
 \begin{align}
 \begin{split}
     \phi^{(g)}_a &= R(g)_a^{\ b}\phi_b,\\
     Q^{(g)}_i &= \tilde{R}(g)_i^{\ j}Q_j.
 \end{split}
 \end{align}
 Here, $R$ and $\tilde{R}$ are representations of the group element $g\in G$. 
 Both $R$ and $\tilde{R}$ are assumed to be non-trivial for the mechanism to work,
 though they are allowed to be reducible and may also contain trivial representations. 
 We consider a symmetry
 such that the ``action'' $S$ remains invariant
 \begin{equation}
     S(\phi^{(g)}, Q^{(g)}) = S(\phi,Q),\label{eq:mech:Sinv}
 \end{equation}
 and the group elements have determinant $1$ such that
 \begin{equation}
     d\phi^{(g)} = d\phi,
 \end{equation}
 which implies that
 \begin{equation}
     \Psi(Q^{(g)}) = \Psi(Q).
 \end{equation}
 Here, it is also implicitly assumed that 
 the integration contour $\pazocal C$ is invariant under the group action.
 However, this is not a general requirement, because, for instance, 
 if the integral \eq{eq:mech:genPsi} is a holomorphic one, like in the case of \eq{eq:review:Psi}, 
$\pazocal C$ is allowed to be transformed up to continuous deformation 
 due to the Cauchy theorem.  
 
 One can make a good estimate of the preferred configurations of $Q$ by considering the critical points $\phi^\sigma$ of the action:
 \begin{equation}
     \left.\frac{\partial S}{\partial\phi_a}\right\vert_{\phi=\phi^\sigma} = 0.\label{eq:mech:critP}
 \end{equation}
 Summing up all the contributions of such critical points for the approximation of the full integral 
 is called the stationary phase approximation\footnote{This terminology is for real $S$. The analogous method for complex cases   
is given by taking the main orders~\cite{Howls2271} in the Picard-Lefschetz theory~\cite{Witten:2010cx}.}.
 Here, we want to use it to qualitatively predict the most important configurations 
 by analyzing these critical points.
 Usually if there exist several critical points, 
 the contributions of the critical points will have uncorrelated phases
 and will cancel each other out so that the sum will generally be small. 
 However, in some cases it is possible to get a large sum, i.e., if we take a certain configuration $Q_H$ which is invariant under a subgroup $H\subset G$. In this case, 
 because of the invariance of \eqref{eq:mech:Sinv}, the critical points form invariant sets with the same phase 
 along the trajectories of the group action $h\in H$,
 \begin{equation}
     S(\phi^\sigma, Q_H) = S(\phi^{\sigma,(h)}, Q_H^{(h)}) = S(R(h)_a{}^b \phi_b^\sigma, Q_H).
 \end{equation}
Because the phase is constant, all the critical points contained in 
such an invariant set contribute coherently to $\Psi$.\footnote{Since the whole system of $\phi$ is invariant under $H$ for $Q=Q_H$, the coherence is an exact
phenomenon beyond the stationary phase approximation.}
If a critical point $\phi^\sigma$ is contained in the trivial part of the representation $R$, 
 the invariant set contains only one element, and the critical point will be isolated in most cases. 
 On the other hand, if the group action generates a non-trivial set of critical points (a continuous set
 if $H$ is a continuous group), these critical points will give larger contributions than the isolated cases.
  
 Generally speaking, this mechanism would prefer a larger representation space of $H$ 
 in the space of $\phi$ for larger amplification of $\Psi$.
 On the other hand, it seems rather difficult to determine whether higher dimensional 
 symmetries are preferred or not. 
A higher dimensional symmetry will in general form a higher dimensional subspace of the 
representation in the space of $\phi$, and therefore the amplification for one particular
 $Q_H$ will become larger. 
 However, higher symmetries will require more conditions on $Q_H$,
and the net probability of getting higher dimensional symmetries can become smaller.
Therefore, it seems a non-trivial question what kinds of symmetric configurations 
have the largest contributions in the end.
 
 In the previous paper a certain toy model was analyzed, which is very similar to the current model of interest, \eqref{eq:review:Psi}, being obtained by setting $\tphi=1$ in \eqref{eq:review:Psi}:
 \begin{equation}
     \Psi(P) = \int d\phi\ e^{i (\phi^2 + P \phi^3)}.
     \label{eq:previousmodel}
 \end{equation}
 This model was found to indeed support this highlighting mechanism of symmetric configurations.
\section{Highlighted symmetries in the canonical tensor model} \label{sec:general}
In this section, we first analyze the critical points in the action of the integral expression of
the wave function \eq{eq:review:Psi} to obtain the potential peaks expected from
the highlighting mechanism of symmetries discussed in Section~\ref{sec:mechanism}.
We find that $\lambda>0$ is required for the existence of interesting critical points.
We give a numerical method for the explicit evaluation of 
the wave function \eq{eq:review:Psi}.
The regularization and its vanishing limit by means of a contour deformation 
are discussed in some detail.  
Finally we show that, for a few concrete examples, 
we indeed find strong peaks at symmetric configurations in accordance with the
highlighting mechanism.

\subsection{Critical points}
\label{sec:generalargument}
 The `action' of the wave function given in \eqref{eq:review:Psi},
 \begin{equation}
     S(P,\phi,\tilde \phi) = P \phi^3 + \phi^2 \tphi - \frac{4}{27 \lambda} \tphi^3,\label{eq:gen:action}
 \end{equation}
 is invariant under the following $O(N)$ transformations,
 \begin{align}
 \begin{split}
     P_{abc} &\rightarrow  L_{aa'}L_{bb'}L_{cc'} P_{a'b'c'},\\
     \phi_a &\rightarrow  L_{aa'} \phi_{a'},
     \end{split}
     \label{eq:underlyingsym}
 \end{align}
 where $L$ denotes the fundamental $O(N)$ matrices.
The existence of such an underlying symmetry is one of the conditions for the mechanism 
in Section~\ref{sec:mechanism} to work, 
and the present model may have the highlighting phenomena for the symmetric configurations.

To start, let us analyze the critical points of \eq{eq:gen:action}. 
From \eqref{eq:mech:critP} one finds the equations to be given by 
 \begin{align}
     \begin{split}
         3 (P \phi^{\sigma\,2})_a + 2 \phi^\sigma_a \tphi^\sigma &= 0,\\
         \phi^{\sigma\,2} - \frac{4}{9 \lambda}  \tphi^{\sigma\,2} &= 0,
     \end{split}
     \label{eq:criticalequations}
 \end{align}
 where  $\phi^\sigma$ and $\tilde\phi^\sigma$ denote the critical points
 of $\phi$ and $\tilde \phi$ respectively, labeled by $\sigma$.
If $\tphi^\sigma\neq 0$ we can rewrite the equations with the rescaled variable
$ \varphi^\sigma=\phi^\sigma/ \tphi^\sigma$, and obtain
 \[
  \begin{split}
         3 (P \varphi^{\sigma\,2})_a + 2 \varphi_a^\sigma&= 0,\\
          \varphi^{\sigma\,2} - \frac{4}{9 \lambda} &= 0.
     \end{split}
 \]
The first equation is nothing but the critical point equation of the previous model \eq{eq:previousmodel}. Therefore, one can expect the symmetry highlighting phenomena will
similarly occur in the present model as the previous one \eq{eq:previousmodel} (as analyzed in~\cite{Obster:2017pdq}). 
Here the only difference comes from the second equation, which restricts the 
size of $\varphi^\sigma$ to be a constant. Since, from the first equation, the size of $\varphi^\sigma$
is inversely proportional to that of $P$, the second equation actually restricts the size of $P$. 
This implies that, unlike the previous model \eq{eq:previousmodel}, 
the values of $P$ for which there exist critical points are restricted by an additional condition.  
This can be obtained by deleting $\tilde \phi^\sigma$ from \eq{eq:criticalequations}:  
 \begin{equation}
     \lambda = \frac{(P \phi^{\sigma\, 3})^2}{(\phi^{\sigma\, 2})^3}.\label{eq:gen:lambda_relation}
 \end{equation}
This is indeed independent of the overall scale of $\phi^\sigma$, giving a restriction on
 the size of $P$. 
 This relation is consistent only for $\lambda>0$. For $\lambda<0$, from \eq{eq:criticalequations},
 there are no other critical points than the trivial one $\phi=\tphi=0$, 
 and we cannot expect for the wave function to have interesting 
structures.\footnote{As a check of this statement, we have computed the wave function for 
some cases with $\lambda<0$ by the method explained in the later sections, 
and have actually observed the monotonous nature of the wave function.} 
Because $\lambda=0$ has already been discarded in Section~\ref{sec:review}, we will exclusively consider the $\lambda>0$ case in this paper.
   
Following the argument of Section~\ref{sec:mechanism}, we want to consider the subgroups of the full symmetry group $O(N)$ under which the configuration $P_{abc}$ itself is invariant. These configurations are expected to be more relevant than the configurations without such symmetric properties. The primary possibilities of symmetries 
 are $O(n)\ (n<N)$, $O(n)\times O(m)\ (n+m<N)$, 
and so on. Other interesting possibilities are the Lie-groups with real orthogonal representations,
which can be embedded into the $O(N)$ matrices of the underlying 
symmetry \eq{eq:underlyingsym}.
To see whether these configurations are preferred,  
it is necessary to develop some tools to actually calculate the integral \eqref{eq:review:Psi}. 
This is highly nontrivial, and few things can be done analytically.
But numerical tools will prove to be of value.
 
\subsection{Evaluating the wave function}
\label{sec:method}
 It is possible to reduce the $N+1$ dimensional integral of \eqref{eq:review:Psi} to an $N$ dimensional compact integral. This is valid for generic $P$, and hence is very useful to analyze the behavior at generic $P$. Though the expression is in principle exact, numerical evaluation might be slow and inaccurate in general due to the remaining numerical 
integration over $N$ variables.
 
\begin{figure}
 \begin{center}
 \includegraphics[scale=.35]{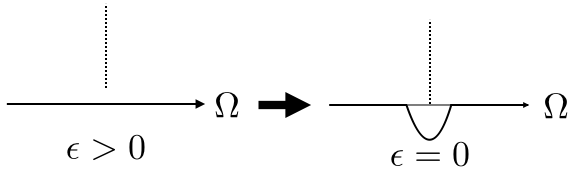}
 \hfil
 \includegraphics[scale=.35]{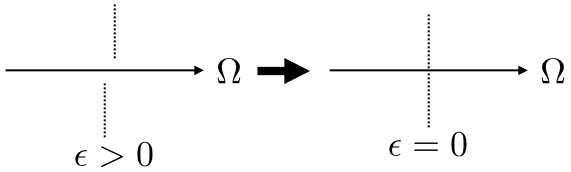}
 \end{center}
 \caption{An illustration of the deformation of the integration contour. 
 The integration contours are described by the thick lines, and the branch cuts of the integrand by the dotted ones. 
 On the left, for $\epsilon>0$, the contour can be taken on a real line, and 
 can safely be deformed without being singular in the $\epsilon\rightarrow 0^+$ limit.
The deformation cannot be done, if the integration contour is pinched by two 
 branch cuts in the $\epsilon\rightarrow +0$ limit, as in the right figure.}
 \label{fig:epsdeform}
 \end{figure}

Taking advantage of the fact that $S(P,\phi,\tilde \phi)$ in \eq{eq:gen:action} is 
a homogeneous cubic function of $\phi$ and $\tilde \phi$, 
we can do a coordinate transformation to hyperspherical coordinates. 
For notational simplicity, we introduce $\tilde{P}$ as the tensor expressing 
$\tilde P_{abc} \phi_a \phi_b \phi_c=S(P,\phi,\tilde \phi)$ in \eq{eq:gen:action}, 
where we regard $\phi_{N+1}=\tilde \phi$.
With hyperspherical coordinates, $\phi$ can be decomposed into an angular and a radial part 
as $\phi_a=\phi_{\Omega\,a} r$, where $\Omega$ denotes a set of angular coordinates,
and $\phi_\Omega$ is a unit vector oriented in the direction described by $\Omega$. 
By this change of variables, we obtain
 \begin{equation}
     \Psi(P) = \int_{S^N} d\Omega \int_0^\infty dr\ r^N e^{i (\tilde P \phi_\Omega^3) r^3 - \epsilon\, r^3},
     \label{eq:gen:psiexpress}
 \end{equation}
 where we have introduced the $\epsilon\, r^3$ term with a positive small $\epsilon$ as a regulator. The $r$ integration can now be done by using $\int_0^\infty dr\ r^N e^{-(\epsilon-i a) r^3} = \frac{1}{3}\Gamma(\frac{1+N}{3})(\epsilon-i a)^{-\frac{1+N}{3}}$:
 \begin{equation}
     \Psi(P) = \frac{1}{3}\,\Gamma\left(\frac{1+N}{3}\right)\int_{S^N} d\Omega\ \frac{1}{(\epsilon 
     - i \tilde P\phi_\Omega^3)^{\frac{N+1}{3}}}.
     \label{eq:wavefunctionomega}
 \end{equation}
Here the branch cut of the fractional power is assumed to be taken on the negative real axis.
This will always be assumed for other fractional powers and logarithmic functions 
appearing in this paper without further notice.   
Now the problem of computing the wave function is reduced to a compact $N$-dimensional integration.
 However, the $\epsilon\rightarrow 0^+$ limit has an apparent difficulty of diverging on the points 
 $\Omega$ satisfying $\tilde P\phi_\Omega^3=0$.
For generic $\tilde P$, this divergence is not the real property of the wave function, because
the singular points can be circumvented
by considering appropriate deformation of the integration contour away 
from the real plane as shown in 
Figure~\ref{fig:epsdeform}.  Such deformation is allowed because of the Cauchy theorem.
For the numerical computation, one can systematically perform this deformation by 
doing a change of variables which adds some imaginary values to the spherical coordinates:
  \begin{equation}
     \Omega'_j = \Omega_j + i \Delta \frac{\partial( \tilde P \phi_\Omega^3)}{\partial \Omega_j},
     \label{eq:wavefunctioncontour}
 \end{equation}
 where $\Delta$ is a small positive number, 
 $\Omega_j\ (j=1,2,\cdots,N)$ are the spherical coordinates taking real values, and $\Omega'_j$ are
 the deformed coordinates which are generally complex. 
 If $\Delta$ is positive and small enough, this deformation will deform the contour appropriately,
 because\footnote{Here, we implicitly take the expression of 
 $\tilde P\phi^3_{\Omega}$ in terms of a holomorphic function of $\Omega$.}  
 \[
 \tilde P \phi_{\Omega'}^3= \tilde P \phi_\Omega^3 + i \sum_{j=1}^N
 \Delta \left( \frac{\partial( \tilde P \phi_\Omega^3)}{\partial \Omega_j}\right)^2+{\pazocal O}(\Delta^2),
 \label{eq:deformofpphi3}
 \] 
 and the second term on the righthand side 
 effectively add a positive contribution to $\epsilon$
 in the denominator of \eq{eq:wavefunctionomega}.
 This will enable one to smoothly take the $\epsilon\rightarrow 0^+$ limit. 
 With this change of variables, 
 \eq{eq:gen:psiexpress} has now been transformed to
 \begin{equation}
     \Psi(P) = \frac{1}{3}\, \Gamma\left(\frac{1+N}{3}\right )\int_{S^N}
     d\Omega \left| \frac{\partial \Omega'}{\partial \Omega} \right|
    \frac{1}{(- i \tilde P\phi_{\Omega'}^3)^{\frac{N+1}{3}}},
     \label{eq:gen:regwavefunction}
 \end{equation}
which does not contain $\epsilon$ anymore. 
 
 As can easily be seen in \eq{eq:deformofpphi3}, the aforementioned method of deforming the contour does not work properly
 on the $\Omega$ for which $\tilde P\phi_\Omega^3=0$ and $\partial (\tilde P\phi_\Omega^3)/\partial \Omega_i=0\ \forall i$.
 Since the former can be regarded as the derivative of $\tilde P \phi^3$ with respect to $r$ and the latter with respect to the angular variables,
 the condition is nothing but $\partial (\tilde P\phi^3)/\partial \phi_a=0\ \forall a$,
 namely the condition \eqref{eq:mech:critP} for a critical point of $\tilde P\phi^3$. 
 As explained in Section~\ref{sec:generalargument}, 
 non-trivial critical points exist\footnote{Note that the trivial critical point $\phi=0$ plays no role in
\eq{eq:gen:regwavefunction}.} 
 only for special values of $\tilde P$, namely those satisfying \eq{eq:gen:lambda_relation}. 
 In other words, there is a correspondence between the singularities of the wave function and the 
 condition \eq{eq:gen:lambda_relation} derived for the existence of critical points. 
Thus, the expression \eq{eq:gen:regwavefunction} is non-singular and valid 
for generic $\tilde P$, except for special $\tilde P$ allowing the 
existence of critical points. 
 A typical example of a singular case is illustrated in the right figure of Figure~\ref{fig:epsdeform}, 
 where the deformation cannot be done because of a pinch by two branch cuts, 
 reflecting the singularity of the wave function on this point in the $\epsilon\rightarrow+0$ limit.
 As used above, in this paper, we often use `generic' and `special'
 to describe the singular and non-singular cases of $P$ (and $\tilde P)$ respectively. 
 
 A consistency check of the final expression of the wave function \eq{eq:gen:regwavefunction} can be done by 
 looking at the dependence of the value on the deformation parameter $\Delta$. 
The wave function should not depend on $\Delta$ due to the Cauchy theorem, 
unless the contour crosses some cuts or singularities by the deformation. 
A more thorough consistency check
is to directly see whether the constraint equations \eq{eq:review:CTMwheelerdewitt} and \eq{eq:review:CTMmomentum} are satisfied by \eq{eq:gen:regwavefunction}.
We have obtained some fairly good numerical results supporting the validity of \eq{eq:gen:regwavefunction}. 
The details are given in the last part of Appendix~\ref{app:wavefunction}. 

The expression  \eq{eq:gen:regwavefunction} has the advantage that it is in principle valid for generic $P$ and any $N$. 
It is certainly useful to study the wave function at general values of $P$
for small $N$.
However, it still contains the $N$-variable integration, so for larger $N$ 
the numerical integration takes much time and may contain large numerical errors. 
Therefore it is not really useful for studying
the large-$N$ behavior of the wave function. 
For that reason, in Section~\ref{sec:simplified},
we consider a subspace of $P$, where one can analytically perform
all the integrations but one, which is numerically evaluated. This subspace
is still large enough to contain both the symmetric and non-symmetric configurations,
and is therefore useful for our purpose to see the highlighting phenomena of 
symmetric configurations.

 \subsection{An example}\label{sec:gen:examples}
  To give a concrete example of the application of the mechanism explained 
  in Section~\ref{sec:mechanism} we consider $N=3$. The case of $N=3$ has just one subgroup
  which can be highlighted by the mechanism: 
  $O(2)\subset O(3)$.  A tensor $P_{abc}$ invariant under an $SO(2)$ transformation can be 
  obtained by solving
  \begin{equation}
      T_{aa'} P_{a'bc} + T_{bb'} P_{ab'c} + T_{cc'} P_{abc'} = 0,\label{eq:gen:ex:gen_eq}
  \end{equation}
  where $T_{ab}$ is the generator of the $SO(2)$ transformation. 
  There always exists an $O(3)$ transformation to put this generator in the following form,
  \begin{equation}
      T = \begin{bmatrix}
      0 & 1 & 0\\
      -1 & 0 & 0\\
      0 & 0 & 0
      \end{bmatrix}.
     \label{eq:so2mat}
  \end{equation}
 \eqref{eq:gen:ex:gen_eq} can then be found to be solved by
  \begin{align}
     \begin{split}
         P_{113}&=P_{223} = \frac{x}{3},\\
         P_{333}&=y,\\
     \end{split} 
  \end{align}
  where the other components (up to permutations) are zero. The action of \eqref{eq:gen:action} for this configuration is given by
  \begin{equation}
      S = x(\phi_1^2+\phi_2^2)\phi_3 + y \phi_3^3 + (\phi_1^2+\phi_2^2+\phi_3^2)\tphi-\frac{4}{27 \lambda} \tphi^3,\label{eq:gen:ex:N=3_S_symmetric}
  \end{equation}
  which is $O(2)$ invariant. The (real) critical points \eqref{eq:mech:critP} of the action
 are given by
  \begin{align}
      &\phi_1=\phi_2=\phi_3=\tphi=0, &\ \\
      &\phi_1=\phi_2=0, \ \phi_3 = -\frac{2}{3y}\tphi, &\lambda&=y^2,\label{eq:gen:ex:N=3_critp_1} \\
      &\phi_1^2+\phi_2^2=R^2 \tphi^2,\ \phi_3=-\frac{1}{x}\tphi, &\lambda&=\frac{4 x^3}{27(x-y)},\label{eq:gen:ex:N=3_critp_2}
  \end{align}
 where $R^2=\frac{2x-3y}{x^3}> 0$. 
  The result is really similar to the one found for the previous model 
  \eq{eq:previousmodel} \cite{Obster:2017pdq}, with the major difference coming from \eqref{eq:gen:lambda_relation}: 
  Each critical point, except the trivial one, 
  has a restriction on the configuration variables, $x$ and $y$, which is shown as the second equation in each line. 
    
  \begin{figure}
\begin{center}
    \begin{subfigure}[]{0.35\textwidth}
     \includegraphics[width=.9\linewidth]{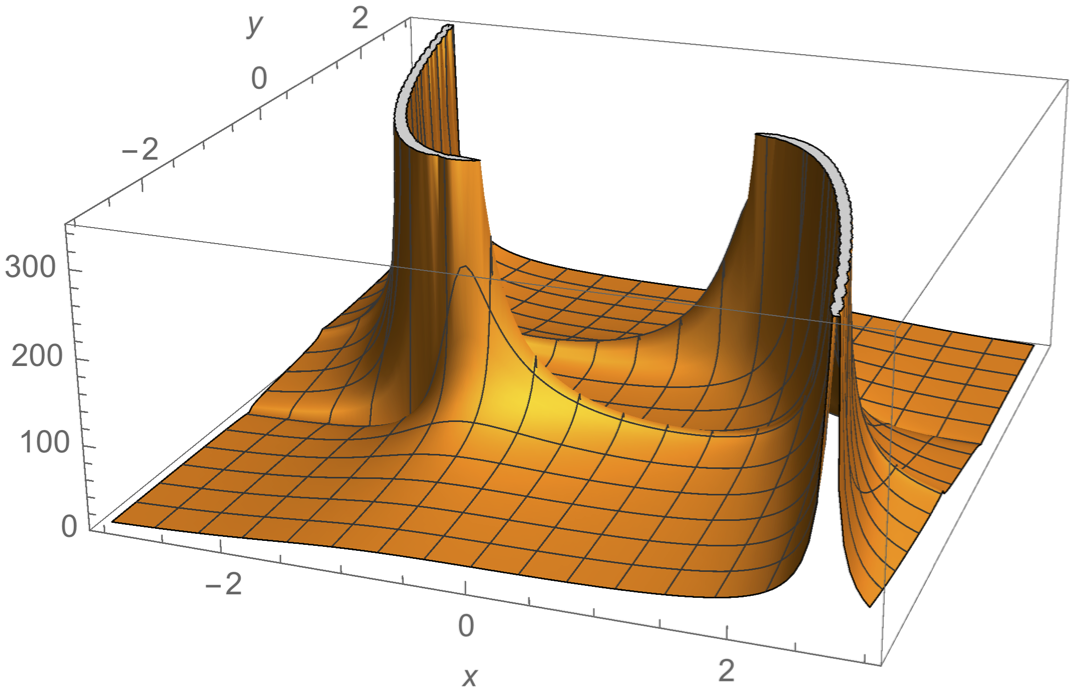}
      \end{subfigure}
    \begin{subfigure}[]{0.3\textwidth}
    \includegraphics[width=.9\linewidth]{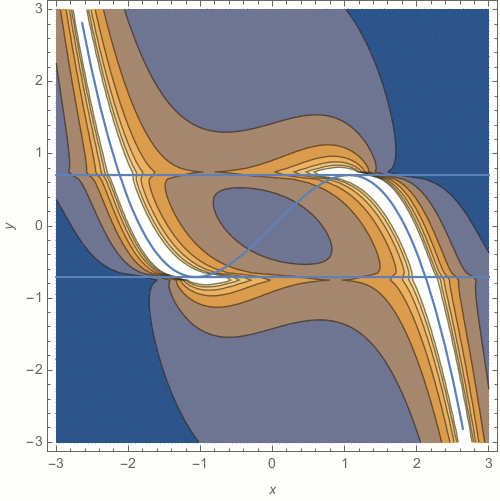}
    \end{subfigure}
    \begin{subfigure}[]{0.3\textwidth}
    \includegraphics[width=.9\linewidth]{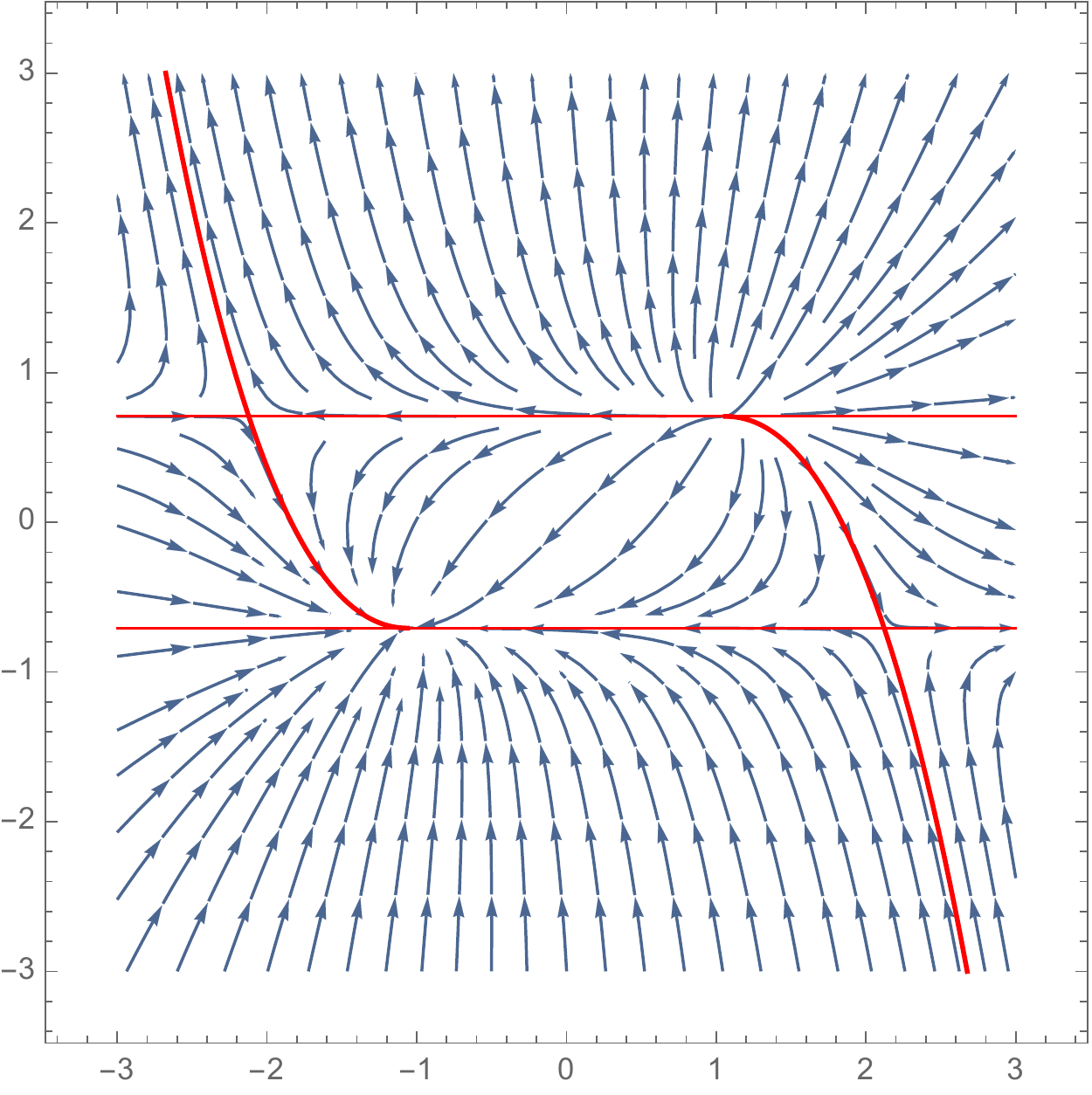}
    \end{subfigure}
    \caption{On the left: A 3D plot of the absolute value of the wave function \eq{eq:gen:regwavefunction} 
    for \eq{eq:gen:ex:N=3_S_symmetric} with $\lambda=0.5$, drawn on the $(x,y)$ parameter plane. 
    The values over 300 have been chopped.
    In the middle: The same function as the left, but as a contour plot.
    The blue solid curve and the straight lines 
    respectively represent the restrictions, $y=x-\frac{4}{27 \lambda} x^3$ in
     \eq{eq:gen:ex:N=3_critp_2} and $y^2=\lambda$ in \eq{eq:gen:ex:N=3_critp_1}. 
     The configurations along \eqref{eq:gen:ex:N=3_critp_2} with $R^2>0$ are strongly peaked. 
     The smaller effects along \eqref{eq:gen:ex:N=3_critp_1} are also visible. 
     On the right: The arrows represent the Hamiltonian vector flow of ${\pazocal H}_3$ of
     the classical CTM.
     }
     \label{fig:gen:ex:P_repr:N=3:SO(2)_contour_xy}
         \end{center}
  \end{figure}

Figure \ref{fig:gen:ex:P_repr:N=3:SO(2)_contour_xy} displays clearly the symmetry 
highlighting phenomenon
  of the present model consistent with the discussions in Section~\ref{sec:mechanism}: The wave function has strong peaks along the trajectory represented by the 
  second equation of  \eqref{eq:gen:ex:N=3_critp_2} with $R^2>0$, while 
  there are small peaks along the second equation of \eqref{eq:gen:ex:N=3_critp_1}.
  As discussed there, one can see that the continuous critical points, namely \eqref{eq:gen:ex:N=3_critp_2}, contribute 
  much larger than the isolated one \eqref{eq:gen:ex:N=3_critp_1}.
  One can also observe that the peaks exist along a classical path of the CTM as shown in the right figure,
   where the classical equation of motion is given by
   \[
   \frac{dP_{abc}}{dt}=\left\{ {\pazocal H}_3,P_{abc} \right\}
   \label{eq:flowofh3}
  \]
  with an auxiliary parameter $t$. 
  Here, the Hamiltonian vector flow of ${\pazocal H}_3$ has the directions 
  within the $x,y$ plane, since this preserves the $O(2)$ symmetry. 
  Thus, the strong peaks can be regarded as representing a particular classical path,
  giving an explicit example of the emergence of classicality in the quantum CTM. 
    
 One can add other terms to the action of \eqref{eq:gen:ex:N=3_S_symmetric} to disrupt the symmetry. From the earlier discussion one would expect that any such terms will make the wave function smaller. 
 For this example we will focus on one term specifically; the others can also be calculated well 
 and the asymptotic properties of these cases will be discussed in Section~\ref{sec:asymptotic}. 
 The term we add here is given by
  \begin{equation}
      \delta S = z \phi_1 \phi_3^2,
  \end{equation}
  where $z$ is a parameter corresponding to $P_{133}=z/3$. 
  One can reduce the integral with this term to a single compact integral:
  \begin{align}
      \Psi(P) &= \int d\phi d\tilde\phi \,e^{i (\phi_1^2(x\phi_3 + \tphi) + \phi_2^2(x\phi_3 + \tphi) + z\phi_1 \phi_3^2 + y \phi_3^3 + \phi_3^2 \tphi - \frac{4}{27 \lambda} \tphi^3 )}\nonumber\\
      &=\int d\phi_3 d\tphi \frac{i \pi }{x \phi_3 + \tphi} e^{i (y \phi_3^3 + \phi_3^2 \tphi - \frac{4}{27 \lambda} \tphi^3 - \frac{z^2 \phi_3^4}{4 (x \phi_3 + \tphi)})}\nonumber\\
      &= \int dr d\theta \frac{i \pi}{x \cos{\theta} + \sin{\theta}} e^{i (y \cos^3{\theta} + \cos^2{\theta}\sin{\theta}- \frac{4}{27 \lambda} \tphi^3 -  \frac{z^2 \cos^4{\theta}}{4(x \cos{\theta} + \sin{\theta})}) r^3}\nonumber\\
      &= i \pi \Gamma(4/3) \int d\theta \frac{\left(-i \left(y \cos^3{\theta} + \cos^2{\theta}\sin{\theta} - \frac{4}{27 \lambda} \sin^3{\theta} - \frac{z^2 \cos^4(\theta)}{4(x \cos{\theta} + \sin{\theta})}\right)\right)^{-1/3}}{x \cos{\theta} + \sin{\theta} },\label{eq:gen:ex:p1p32_exact}
  \end{align}
  where a similar deformation of the integration contour as in \eq{eq:gen:regwavefunction} is implicitly assumed.
  
  \begin{figure}
   \centering
   \includegraphics[width=.5\linewidth]{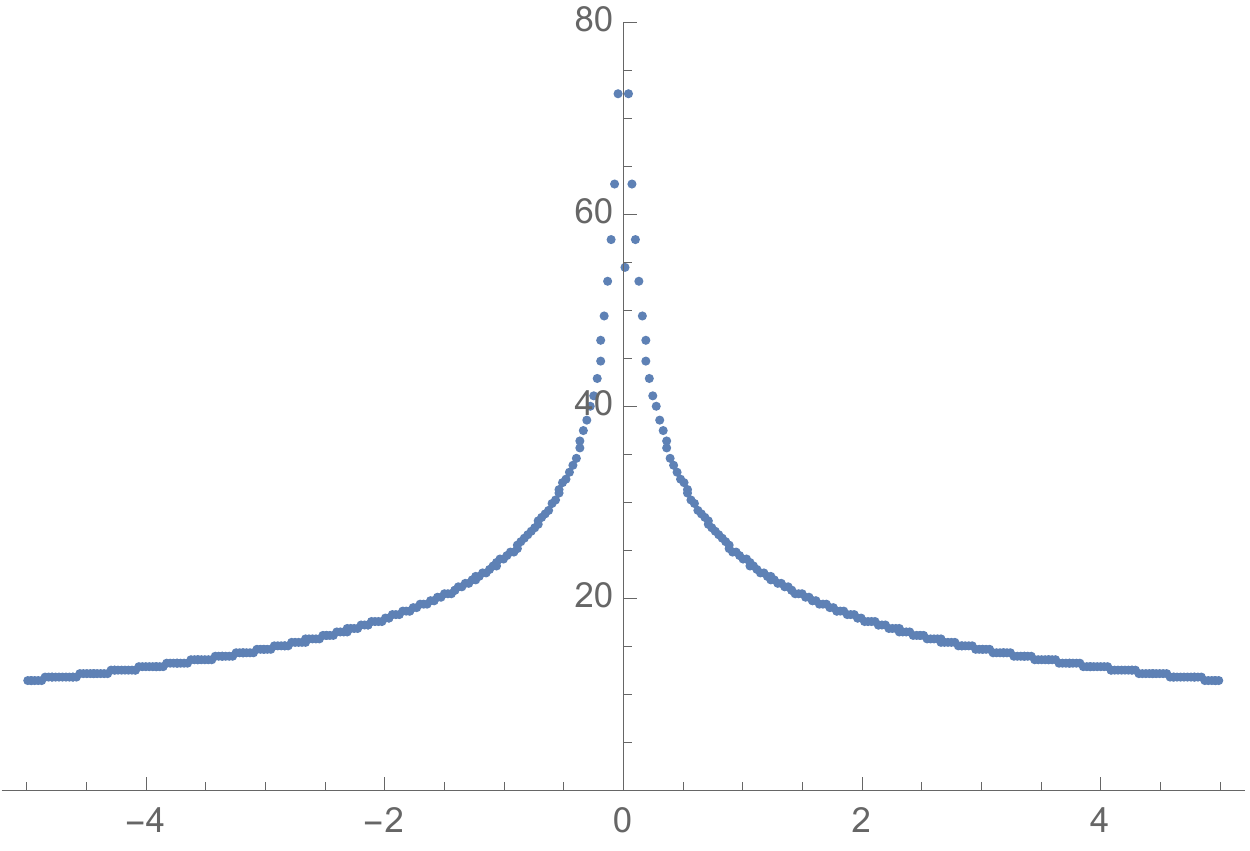}
   \caption{The evaluation of the integral \eqref{eq:gen:ex:p1p32_exact} with $x=3$, $y=x-\frac{4}{27 \lambda} x^3$, $\lambda=1$. This is plotted against $z$, and 
   the symmetric configuration $z=0$ is clearly preferred.}
   \label{fig:gen:ex:p1p32_exact}
  \end{figure}
    
  The integral \eqref{eq:gen:ex:p1p32_exact} can be evaluated numerically, one example of which can be found in Figure~\ref{fig:gen:ex:p1p32_exact}. One can see that indeed the wave function has a peak for $z=0$, so the symmetric configuration is preferred. 
\section{Simplified calculable model}
\label{sec:simplified}
 In this section we treat a simplified model by restricting $P$ to a subspace. The motivation for this is to reduce the number of numerical integrations needed for the evaluation of the wave function,
as the numerical integrations over multiple variables take much time and are not always 
reliable due to numerical errors. In this subspace,
 the integrations in \eq{eq:review:Psi} can be done analytically except for one integration, which 
 can be numerically evaluated much more easily. 
The restriction does not spoil our main purpose: The subspace is large enough to observe 
the symmetry highlighting phenomenon in the CTM, though it is not enough
to predict the most preferred symmetric configuration in the full configuration space.
Because of the single remaining numerical integration, 
the model can be used to numerically study the behavior of the wave function for large $N$,
which is virtually impossible by the method described in Section~\ref{sec:method}.

The model we consider is given by
 \begin{equation}
     \Psi(x,y) = \int d\phi d\tphi\ e^{i S(\phi, \tphi, x, y)},
     \label{eq:simplepsi}
 \end{equation}
 where,
\begin{equation}
     S(\phi, \tphi, x, y) = \sum_{i=1}^{N-1} x_i \phi_i^2 \phi_N + y \phi_N^3 + \phi^2 \tphi - k  \tphi^3
\label{eq:simpleaction}
 \end{equation}
 with $k=\frac{4}{27 \lambda}$.
 When $x_i$ are all equivalent, the action is invariant under an $O(N-1)$ symmetry. When
 all values of $x_i$ are different, the $O(N-1)$ symmetry is broken to $O(1)^{N-1} \cong Z_2^{N-1}$. 
We also have the intermediate cases with 
$\bigotimes_{i=1}^{k} O(n_i)$ with $n_1+...+n_k = N-1$ by considering some sets of equivalent 
$x_i$ as the following illustrative example:
\begin{equation}
 (x_i) = \left(\vphantom{x_1,x_1,x_1,x_2,x_2,x_3,x_4,x_4}\right. \overbrace{x_1, x_1, x_1}^{O(3)},\overbrace{x_2,x_2}^{O(2)},\overbrace{x_3}^{O(1)},\overbrace{x_4,x_4}^{O(2)},... \left. \vphantom{x_1,x_1,x_1,x_2,x_2,x_3,x_4,x_4} \right).
\label{eq:simplified:O(n)_illustration}
\end{equation}
This fact makes this model interesting for the study of the symmetry highlighting phenomena, 
though the subspace is really small compared to the whole space of $P$. 

One can analytically integrate over $\phi_i\ (i=1,2,\cdots,N-1)$
in \eq{eq:simplepsi}, because these are just Gaussian integrations.  
After this exercise there remain two integrations, over $\phi_N$ and $\tilde \phi$, and one of them
can be done in a similar manner as the radial integration in \eq{eq:gen:psiexpress}.
By doing this integration we obtain
 \begin{equation}
     \Psi=2 \pi^\frac{N-1}{2} \hbox{Re}\left[ A_+\right],
 \end{equation}
 where $A_+$ is given by
 \begin{equation}
    A_+=\frac{1}{3} \Gamma\left( \frac{5-N}{6}\right) \int_{\pazocal C} d \tilde \phi
\prod_{j=1}^{N-1} \frac{1}{\sqrt{ - i (x_j+\tilde \phi)}} \left(-i h(\tilde \phi)\right)^\frac{N-5}{6},
\label{eq:simpleap}
 \end{equation}
with $h(\tilde \phi)=\tilde \phi+y  - k\, \tilde \phi^3$ if $(N-5)/6$ is not an integer, or \eqref{eq:apint} if $(N-5)/6$ is an integer.
 The detailed derivation is given in Appendix~\ref{app:reduce}. The integration contour $\pazocal C$ 
 should be taken so as to avoid the branch cuts of the integrand
  as illustrated in Figure \ref{fig:contourtilde}.
This is determined by the vanishing limit of the regularization, as explained in Appendix~\ref{app:reduce}. 
Since there remains only one integration, the numerical evaluation is relatively easy, even for a large $N$.
 \begin{figure}
\begin{center}
\includegraphics[scale=.3]{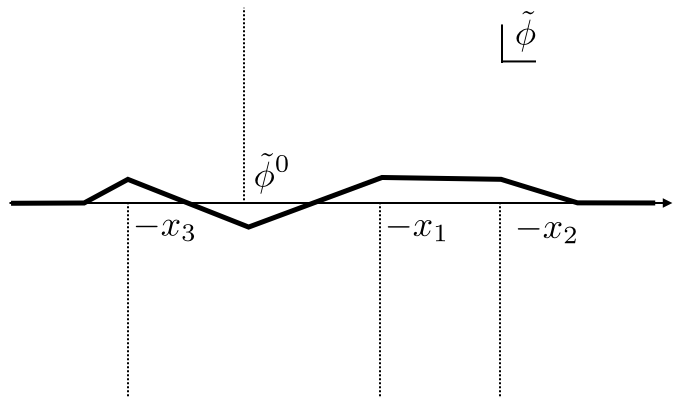}
\caption{An example of the integration contour ${\pazocal C}$ of $\tilde \phi$. 
The dotted lines are the branch cuts of the integrand. $\tilde \phi^0$ denotes 
a real solution to $h(\tilde \phi)=0$. The example is shown for $h'(\tilde \phi^0)<0$.}
\label{fig:contourtilde}
\end{center}
\end{figure}

From expression \eq{eq:simpleap} (or \eqref{eq:apint}), one can easily 
see the symmetry highlighting phenomena in this simplified case. 
As in the right figure of Figure~\ref{fig:epsdeform},
the wave function will have singularities, if the integration contour is pinched by 
some branch cuts. In the example of Figure~\ref{fig:contourtilde}, this occurs 
if some of $-x_i$ coincides with $\tilde \phi^0$, and the number, say $n$, of 
$-x_i$ which accumulates at $\tilde \phi^0$ will determine the highlighted symmetry to
be $O(n)$.

In general, as explained in Appendix~\ref{app:reduce}, the branch cuts of the square roots
of the integrand extends to the negative imaginary direction, while 
those of $(-i h(\tilde \phi))^{(N-5)/6}$ (or $\log(-i h(\tilde \phi))$ for integer $(N-5)/6$ and similarly below) 
extends from the point where $h(\tilde \phi)=0$ on the real axis
to the positive imaginary direction if $h'(\tilde \phi)<0$ at the point and the negative
imaginary direction if $h'(\tphi) > 0$.
Due to the simple cubic form of $h(\tilde \phi)$ in 
$\tilde \phi$, 
there only exist two major cases for the branch cuts of $(-i h(\tilde \phi))^{(N-5)/6}$.
These are 
(i) $y^2>\lambda$: One branch cut extends to the positive imaginary direction, or (ii)
$y^2<\lambda$:
Two brach cuts extend to the positive imaginary direction, 
and one in-between extends to the negative imaginary direction.
Since the pinching occurs only when the positive and negative branch cuts meet,
we have the following three major kinds of singularities of the wave function.
\begin{itemize}
\item At $y^2=\lambda$, two of the branch cuts of $(-i h(\tilde \phi))^{(N-5)/6}$, one extending in the negative imaginary direction and the other in the positive, pinch the contour.
\item At $y^2>\lambda$, $O(n)$ symmetric configurations are highlighted by the 
accumulation of $n$ of $-x_i$ to the one branch point of $(-i h(\tilde \phi))^{(N-5)/6}$.
\item At $y^2<\lambda$, $O(n_1)\times O(n_2)$ symmetric configurations are highlighted
by the similar accumulation above to the two branch points of $(-i h(\tilde \phi))^{(N-5)/6}$ whose 
branch cuts extend to the positive imaginary directions.
\end{itemize} 
The first one corresponds to the small peak described in \eq{eq:gen:ex:N=3_critp_1} for $N=3$,
and the second and third to that in \eq{eq:gen:ex:N=3_critp_2}, where only an accumulation to
one branch point is considered  in \eq{eq:gen:ex:N=3_critp_2} due to $x_1=x_2=x$. 
One important thing to notice is that, while the action \eq{eq:simpleaction}
can have various $\bigotimes_{i=1}^{k} O(n_i)$ symmetries, 
the actual highlighted symmetries are limited to $O(n)$ or $O(n_1)\times O(n_2)$.

\begin{figure}
\begin{center}
\includegraphics[scale=.3]{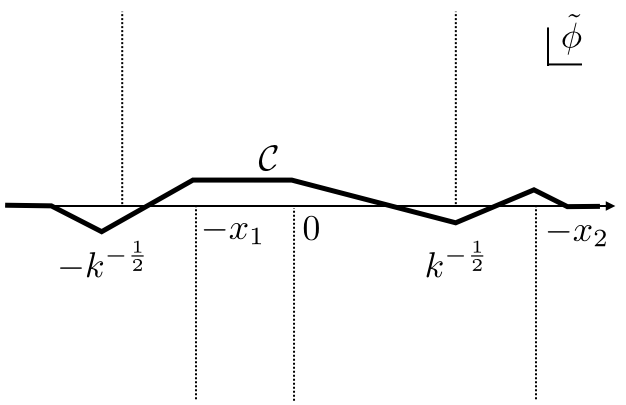}
\hspace{.3cm}
\includegraphics[scale=.3]{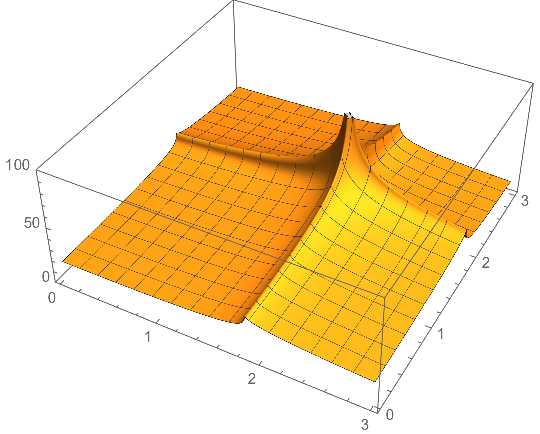}
\end{center}
\caption{
The left: The branch cuts of the integrand and the integration contour ${\pazocal C}$ for \eq{eq:exap}.
The right: A 3D plot of the wave function plotted on the $(x_1,x_2)$ plane for $N=3$, $y=0$ and $\lambda=0.5$.
There are small peaks corresponding to the case that one of the $x_i$ is on $k^{-1/2}$, and a large peak corresponding 
to the highlighting of an $O(2)$ symmetry at $x_1=x_2=k^{-1/2}$. The values over 100 have been chopped.
}
\label{fig:branchcuts}
\end{figure}

As a concrete example, let us consider the simple case with $N=3$, $y=0$ of \eq{eq:simpleaction}. 
In this case, \eq{eq:simpleap} has the form,
\[
A^{N=3,y=0}_+\propto \int_{\pazocal C} d\tilde \phi\ (-i(x_1+\tilde \phi))^{-\frac{1}{2}} (-i(x_2+\tilde \phi))^{-\frac{1}{2}}(-i\tilde \phi (1+\sqrt{k} \tilde \phi)(1-\sqrt{k} \tilde \phi))^{-\frac{1}{3}}. 
\label{eq:exap}
\] 
The branch cuts of the integrand and the integration contour ${\pazocal C}$ 
are illustrated in the left figure of Figure~\ref{fig:branchcuts}. 
There are three branch points coming from $h(\tilde \phi)=0$, i.e., $\tilde \phi=0,\pm k^{-1/2}$, 
but the singular behaviors are expected only for $x_i=\pm k^{-1/2}$, 
not for $x_i=0$ because of the absence of a pinch. 
If only one of $x_i$ is at $\pm k^{-{1/2}}$,
the singularity of the integrand is not strong enough for the integral to diverge, but there will 
be a rapid change of the value  when $x_i$ passes over $\pm k^{-{1/2}}$, 
because the integration contour gets substantially changed.
If $x_1=x_2=\pm k^{-{1/2}}$, the integral diverges, and the wave function has 
a strong peak there\footnote{In this paper, we will not discuss whether this divergence is 
square integrable over $P$ or not, because, for that, we need the understanding of the 
behavior in the full parameter of $P$, which is out of our present reach.}.   
This indeed corresponds to an $O(2)$ symmetric configuration,
and is interpreted as an occurrence of the symmetry highlighting phenomenon 
(See the right figure of Figure~\ref{fig:branchcuts}). 

As mentioned in the beginning of this section, this simplified setting gives a relatively simple way to do some analysis on the large $N$ behaviour and to explore some of the possible symmetric configurations. To do the analysis of the first problem we will introduce a parameterization of just one parameter $z$, which breaks the $O(N-1)$ symmetry completely  
to $O(1)^{N-1}$, given by
\begin{equation}
x_i(z) = x_0+z \left(\frac{i}{N} - \frac{1}{2}\right).
\end{equation}
This parameterization is chosen such that $x_i\ (i=1,2,\cdots,N-1)$ are distributed 
evenly in the region,
\[
x_i(z) \in \left( x_0-\frac{z}{2},x_0+\frac{z}{2}\right)
\label{eq:regionx}
\]
with a center at $x_0$.
Once one understands the behaviour of this for large $N$, one might be able to get a hint  
about the $N\rightarrow\infty$ limit for the simplified model \eq{eq:simpleaction}. 
However, this parameterization reduces the configuration space even further, 
making the predictions somewhat modest.

\begin{figure}
\begin{center}
    \begin{subfigure}[]{0.475\textwidth}
     \includegraphics[width=.9\linewidth]{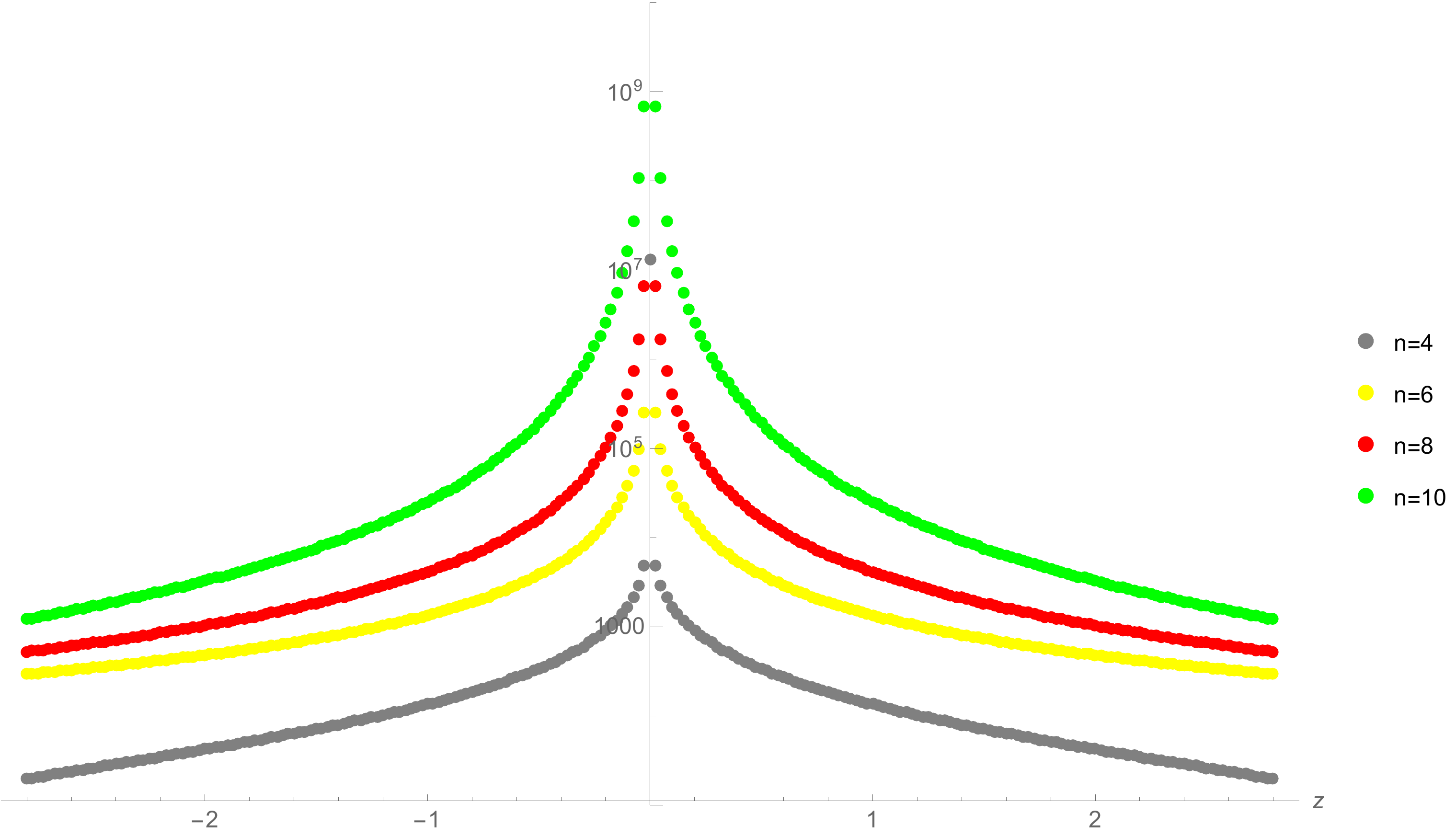}
      \end{subfigure}
     \begin{subfigure}[]{0.475\textwidth}
    \includegraphics[width=.9\linewidth]{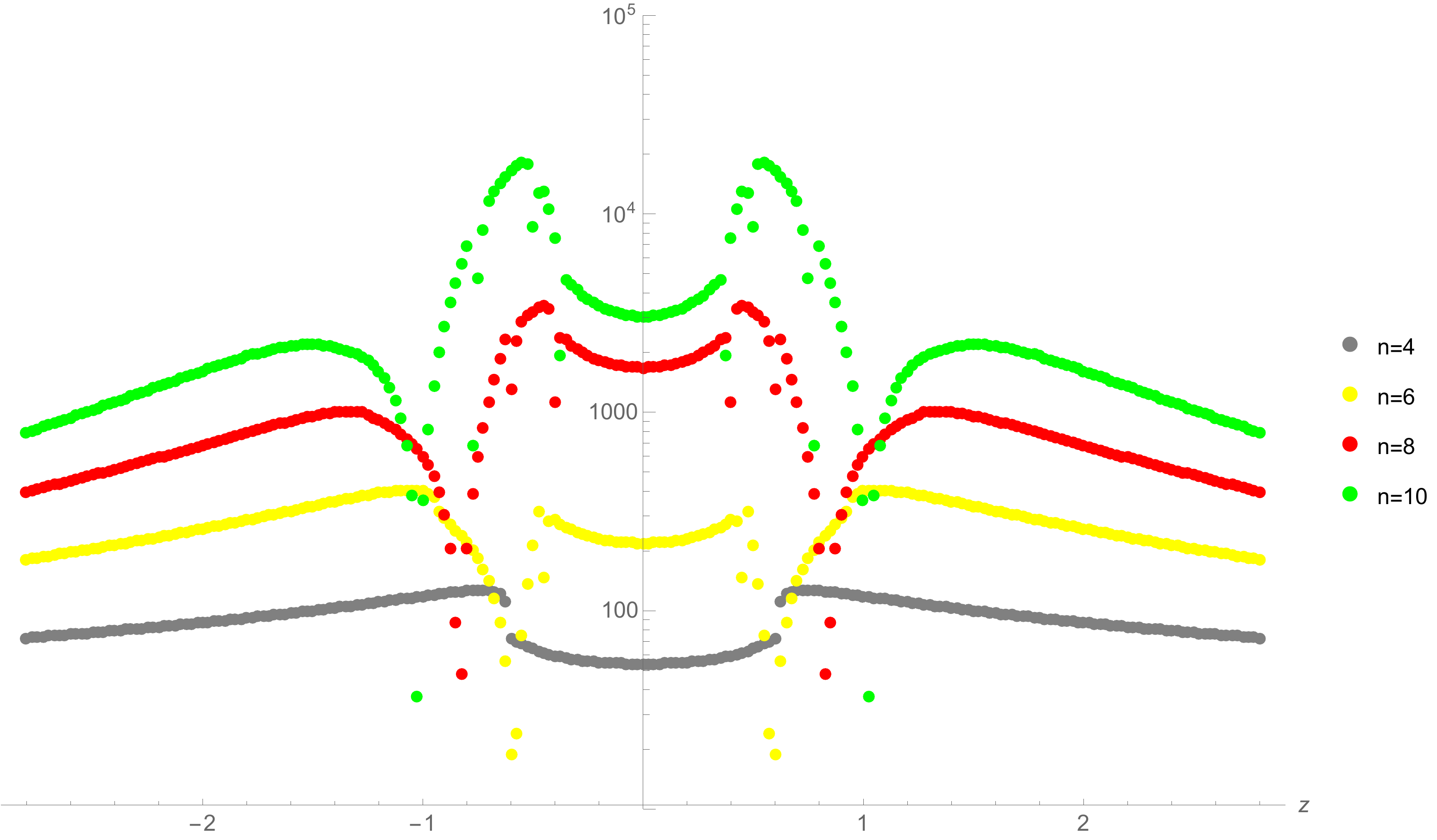}
    \end{subfigure}
    \caption{On the left: The absolute value of the wave function 
    plotted against $z$ for $N=\{4,6,8,10\}$ with $x_0=2$, $y=x_0 - k x_0^3$, $k=4/27$.
     On the right: The same for $x_0=2$, $y=x_0 - k x_0^3 + 0.1$, $k=4/27$. It is clear that  
     even a small number of $\alpha$ changes the overall structure. There even exist the locations where the wave function changes its sign.
These appear as the sharp valleys in the plot of the absolute value. 
       Comparing the two figures, one can see that the peak for $\alpha=0$ is much stronger. }
     \label{fig:simplified:alpha}
         \end{center}
  \end{figure}

For $y$, we shall choose a value of
\begin{equation}
 y = x_0 - k x_0^3 + \alpha,
 \label{eq:yalpha}
\end{equation}
where $\alpha$ is the parameter we will change.
When $\alpha=0$, the relevant branch point of $(-i h(\tilde \phi))^{(N-5)/6}$, written in $x=-\tilde \phi$, 
is located at the center $x=x_0$ of the region \eq{eq:regionx}. Therefore, if we take $|z|$ smaller, 
$x_i$ accumulate toward the branch point and the configuration approaches to 
an $O(N-1)$ symmetric one. Indeed, as seen in the left figure of Figure~\ref{fig:simplified:alpha},
there is a strong peak at $z=0$, as expected from the highlighting mechanism.
The peak is enhanced for larger $N$. 
  
If we take a non-zero value for $\alpha$, $z=0$ is not anymore the case in which 
the accumulation of $x_i$ toward the branch point occurs.
In fact, the region \eq{eq:regionx} contains the branch point only for
 $|z|>z_{min}$ with a positive value of $z_{min}$. 
 The minimum $z_{min}$ can be obtained by solving the condition 
 for one of the endpoints\footnote{Here we ignore a small difference coming from the fact that the endpoints of the region \eq{eq:regionx} are not contained as $x_i$. For large $N$ this is justified.} 
 of the region \eq{eq:regionx} to coincide with the branch point, e.g. 
 $y=x_0+z/2-k (x_0+z/2)^3$ with \eq{eq:yalpha} for $x_0,z>0$.
Indeed, as can be seen in the right figure of Figure~\ref{fig:simplified:alpha},
the peaks are located away from $z=0$. What occurs around the peaks is that
$x_i(z)$ pass over the the branch point 
one by one in the course of changing the value of $z$. 
The highlighted symmetry is just $O(1)$ for each passing over. 
The wave function has some rich structures, seemingly reflecting the one-by-one passing over. 
The amplitude of the wave function is enhanced for larger $N$, but is
substantially smaller than the $O(N-1)$ case in the left figure.
Note also that for non-zero $\alpha$, even though the point $z=0$ still corresponds to 
an $O(N-1)$ symmetric configuration, this is not a large peak, 
since it misses the condition \eq{eq:gen:lambda_relation}. 
Rather, the condition is satisfied for a highlighted $O(1)$ symmetry, 
when one of the $x_i(z)$ passes over the branch point.

\begin{figure}
\begin{center}
    \begin{subfigure}[]{0.475\textwidth}
     \includegraphics[width=.9\linewidth]{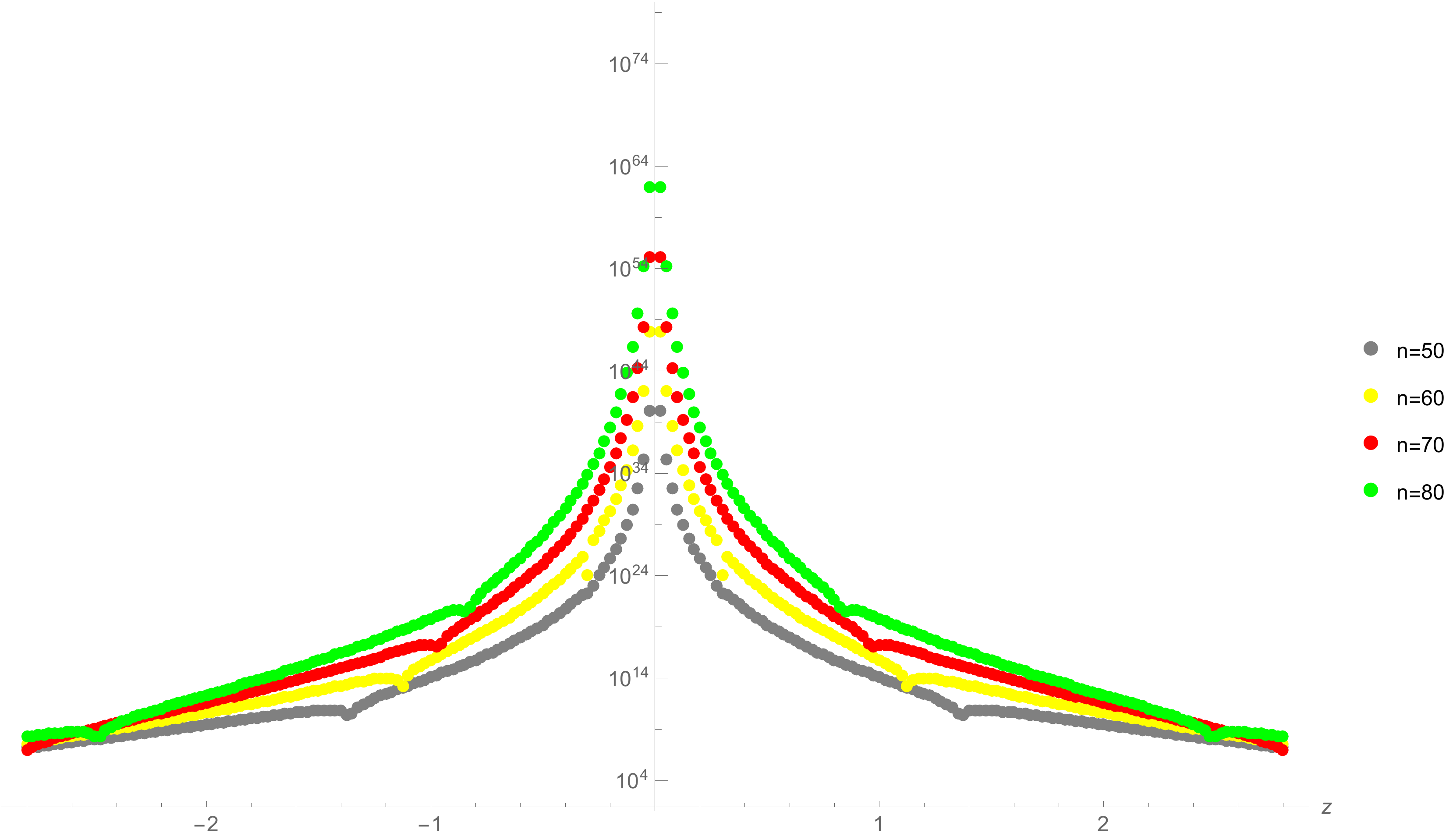}
      \end{subfigure}
     \begin{subfigure}[]{0.475\textwidth}
    \includegraphics[width=.9\linewidth]{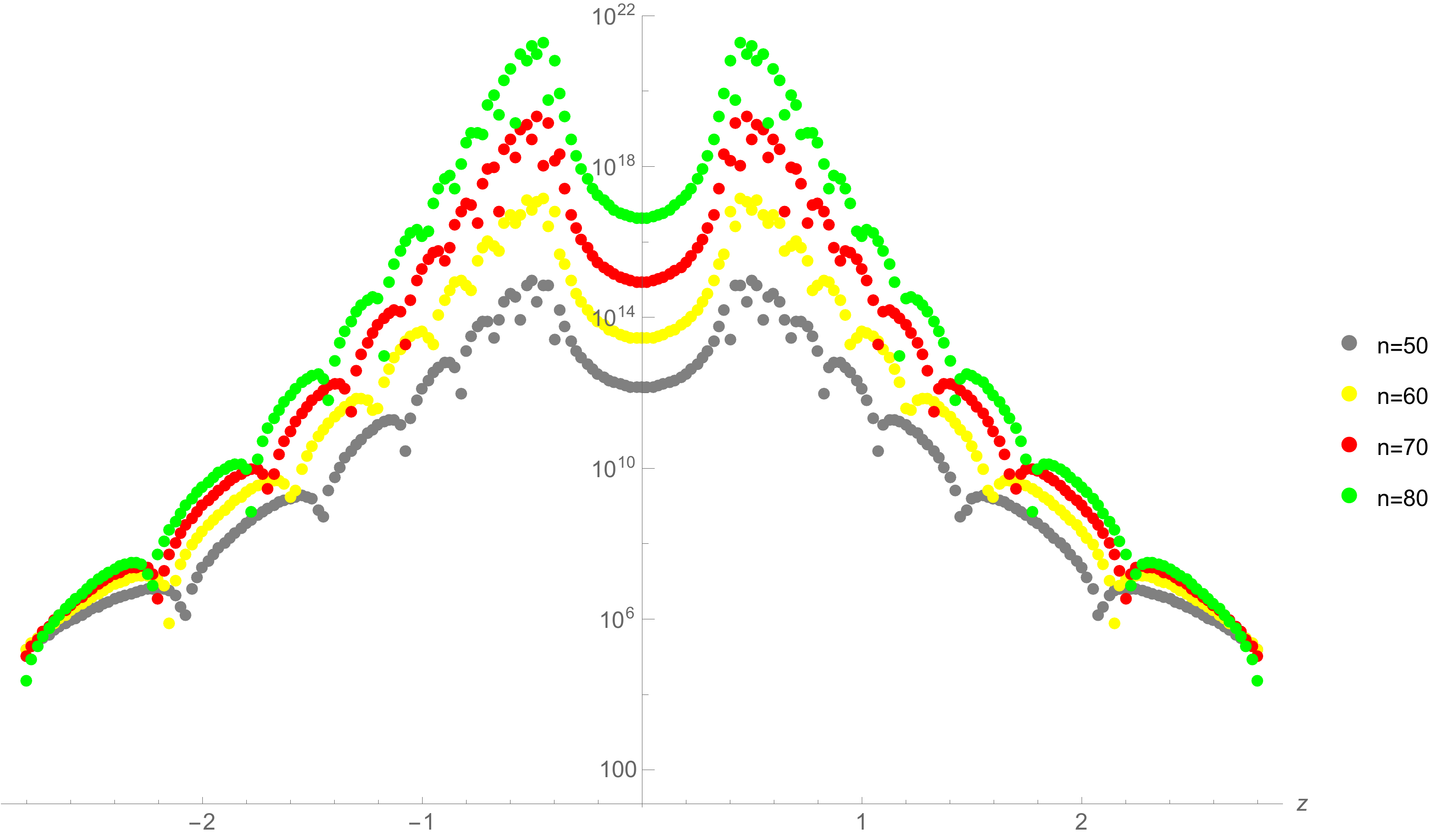}
    \end{subfigure}
    \caption{On the left: The absolute value of the wave function 
    plotted against $z$ for $N=\{50,60,70,80\}$ 
    with $x_0=2$, $y=x_0 - k x_0^3$, $k=4/27$.
     On the right: The same for $y=x_0 - k x_0^3 + 0.1$.
     The wave function for finite $\alpha$ has more structures and is not peaked around $z=0$.
     The peak is much stronger for $\alpha=0$.   }
     \label{fig:simplified:alpha_largeN}
         \end{center}
  \end{figure}

For the large $N$ limit one can also use this setup and see if eventually this wave function will converge to something meaningful. While the case for $\alpha \sim 0$ seems to be
rather simple, the situation for non-zero $\alpha$ seems to be much more complex. 
As explained above, the wave function as a function of $z$ is made of a collection of 
the $O(1)$ symmetric peaks, whose amount is $N-1$. 
Among them, the peaks closer to $z=0$ appear to become more and more important for 
larger $N$, which is due to the high density of peaks in this region.
These things can be seen in Figure~\ref{fig:simplified:alpha_largeN}.

Lastly we will investigate the behaviour of the simplified model for other symmetries. As mentioned at \eqref{eq:simplified:O(n)_illustration}, it is possible to find several subgroups in the configuration $x_i$. Let us consider the case symmetric under a product group, $O(n_1)\times O(n_2)$
with $n_1+n_2=N-1$, having the following $x_i$: 
\begin{equation}
( x_i) = \left(\vphantom{x_1,...,x_1,x_2,...,x_2}\right. \overbrace{x_1, ..., x_1}^{n_1\text{ times}},\overbrace{x_2,...,x_2}^{n_2\text{\ times}} \left. \vphantom{x_1,...,x_1,x_2,...,x_2} \right).
\label{eq:setupx1x2}
\end{equation}
This is supposed to be the maximum possibility of the highlighted product group symmetry, 
because a product group with more than two $O(n_i)$ cannot be highlighted,
as mentioned earlier in the analysis of the singularity of the wave function.
For \eq{eq:setupx1x2}, one can derive the following two sets of continuous critical points
in the same manner as deriving \eq{eq:gen:ex:N=3_critp_2} of the previous example:
\begin{align}
\begin{split}
& \phi_1^2+...+\phi_{n_1}^2 = R_1^2 \tphi^2,\ \ \phi_N = -\frac{1}{x_1} \tphi, \ \hbox{others}=0, \\
&\phi_{n_1+1}^2+...+\phi_{N-1}^2 = R_2^2 \tphi^2,\ \ \phi_N = -\frac{1}{x_2}
  \tphi, \ \hbox{others}=0,\label{eq:simplified:crit_p_OxO}
\end{split}
\end{align}
where $R_i^2=\frac{2x_i-3y}{x_i^3}$.
Each set of the continuous critical points exists only if the extra restriction $y=x_i - k x_i^3$ 
is satisfied as in the second equation of \eq{eq:gen:ex:N=3_critp_2}.
There is another way to derive this restriction by using the singular structure of the branch cuts, 
not by using the stationary phase approximation. As mentioned earlier, 
the wave function has singularities when $h( - x_i)=0$, 
solving this equation will lead to the same retriction. 
Obviously we get the $O(N-1)$ symmetry by putting $x_1=x_2$.
\begin{figure}
\begin{center}
    \begin{subfigure}[]{0.475\textwidth}
     \includegraphics[width=.9\linewidth]{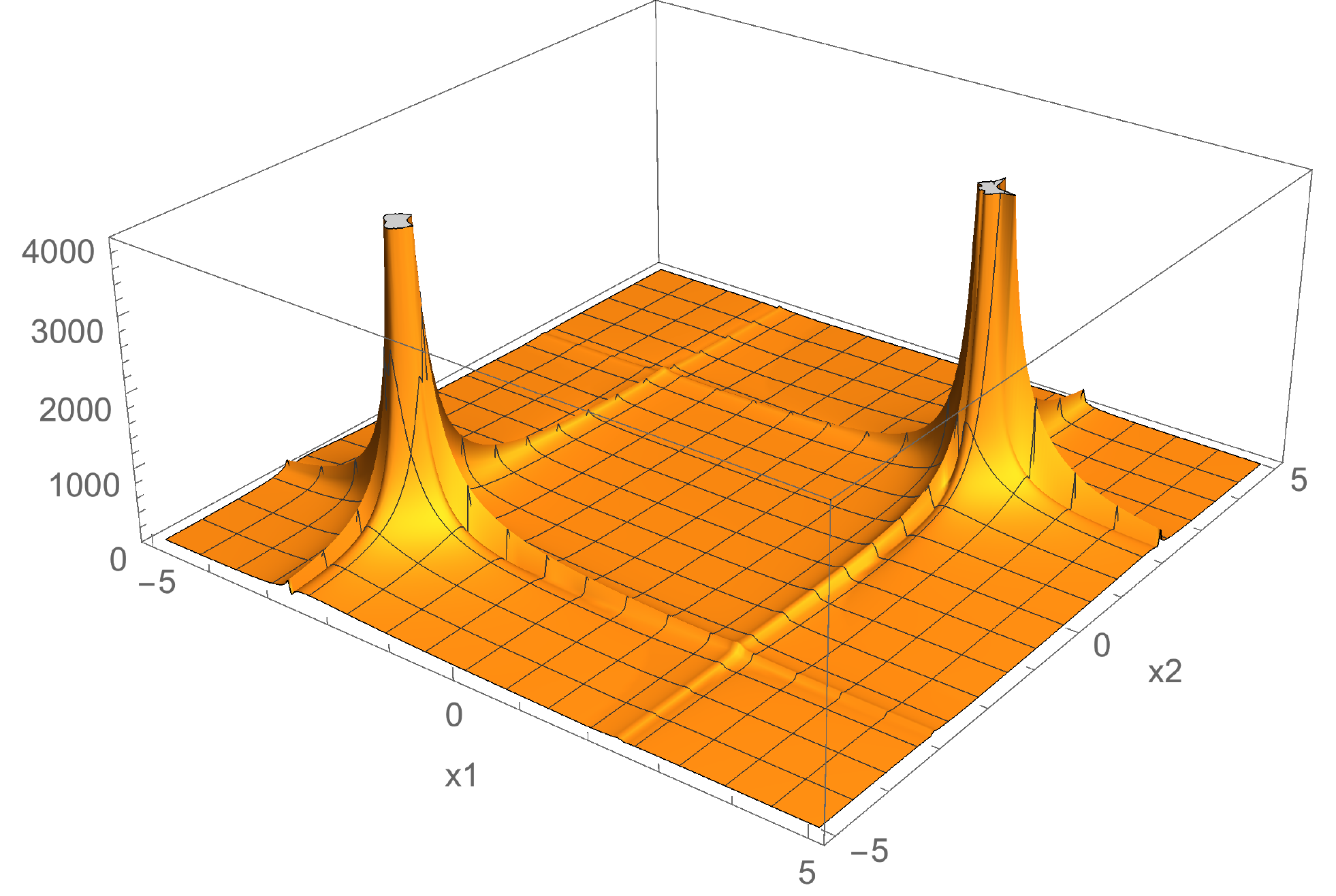}
      \end{subfigure}
     \begin{subfigure}[]{0.475\textwidth}
    \includegraphics[width=.9\linewidth]{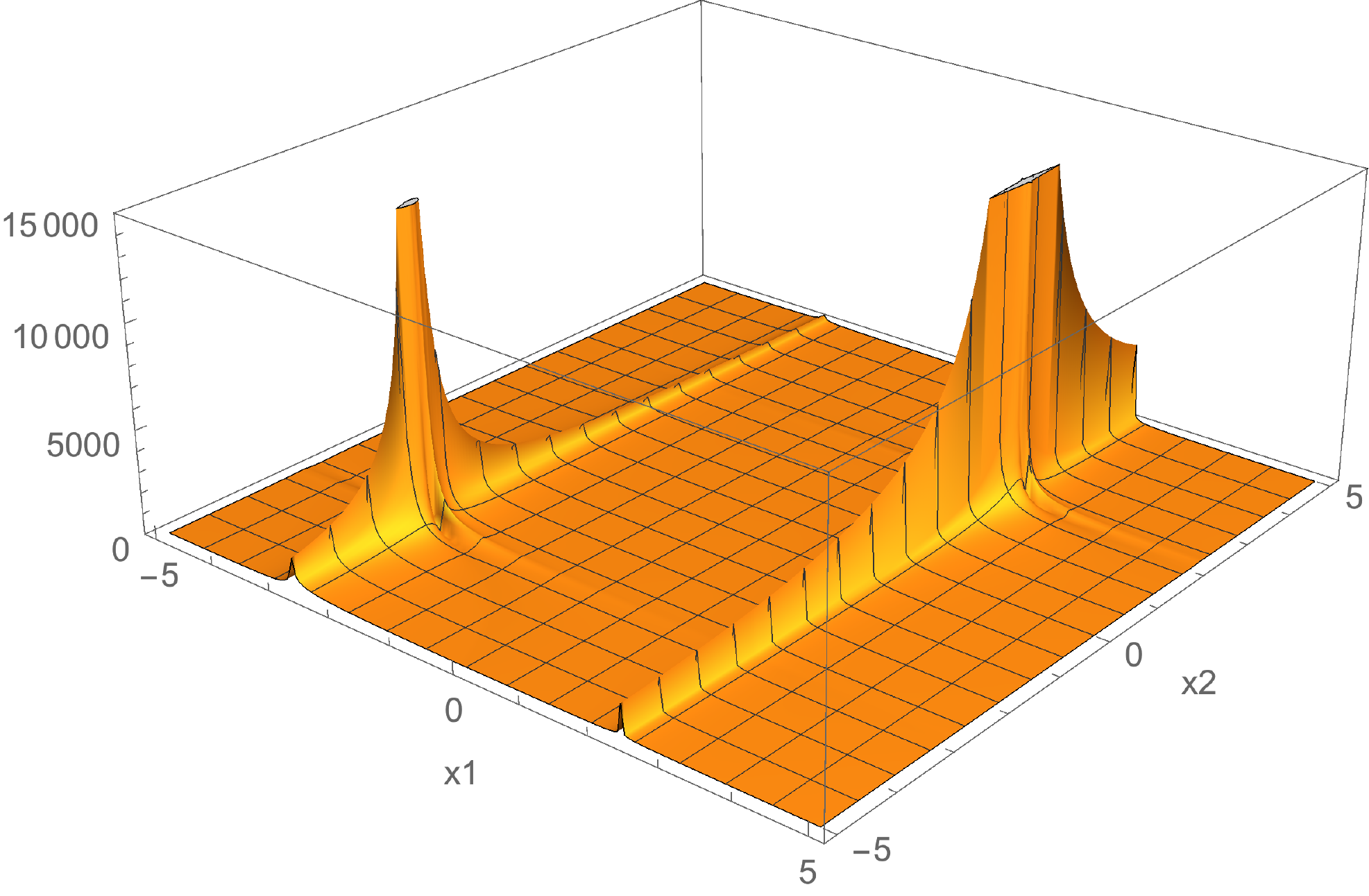}
    \end{subfigure}
    \caption{On the left: The wave function plotted against $x_1,x_2$
    for a setup given by \eqref{eq:setupx1x2} for $O(3)\times O(3)$ symmetry. 
    The largest peaks at $x_1=x_2$ are of the $O(6)$ symmetry, 
    and the shallow peaks extending from them
    the separate $O(3)$ symmetries. 
    At the crossing points of the $O(3)$ peaks corresponding to $O(3)\times O(3)$, 
    one can find a small additive effect. 
     On the right: The same for 
     $O(4)\times O(2)$ symmetry. It can be seen that the $O(6)$ symmetry at $x_1=x_2$ is most preferred, while the $O(4)$ peaks 
     are much stronger then $O(2)$. 
     At the point where $O(4)\times O(2)$ should occur, there should be an interference effect, but this cannot be easily seen because of the large difference of the strengths 
     between $O(4)$ and $O(2)$.
     In both figures, the values over certain ones have been chopped.
     }
     \label{fig:simplified:several_symmetries}
         \end{center}
  \end{figure}

From \eqref{eq:simplified:crit_p_OxO} one would expect two things. Firstly, it seems that the continuous critical points  in \eqref{eq:simplified:crit_p_OxO} have smaller
orbit spaces than $O(n_1+n_2)$, making $O(n_1+n_2)$ probably the stronger symmetry. 
Second, it is possible for the sets of
continuous critical points to coexist 
highlighting the $O(n_1)\times O(n_2)$ symmetry, 
if $y=x_1 - k x_1^3 = x_2 - k x_2^3$ is satisfied with distinct $x_1$ and  $x_2$. 
These two contributions of the continuous critical points interfere with each other,
and they may add up or cancel each other depending on particular cases.
The example for $N=7$ is given in Figure~\ref{fig:simplified:several_symmetries},
where we see the peaks of various patterns of symmetries.  

\section{Emergence of hidden spacetime symmetries}
\label{sec:timelike}
So far, we have only payed attention to the $O(N)$ symmetry,
\eq{eq:review:ON_invariance}, 
which can be regarded as the space-like symmetry of the CTM in analogy with the ADM formalism. 
It would be reasonable to apply the highlighting mechanism to the $O(N)$ symmetry,
since it is the kinematical symmetry of the CTM.
However, while the symmetry is represented on the integration variable $\phi$ of the wave function
\eq{eq:review:Psi}, there exists the other integration variable $\tilde \phi$.
Taking the mechanism more carefully, larger symmetries represented on both $\phi$ and $\tilde \phi$ 
have the possibility to be highlighted. Moreover, we also impose the Hamiltonian constraint
${\pazocal H}_{ab}|\Psi\rangle=0$, not only the kinematical one ${\pazocal J}_{ab}|\Psi\rangle=0$.  
In this section, we will explicitly show that the above prospect for larger
highlighted symmetries is indeed right
in the example discussed in Section~\ref{sec:gen:examples} and 
Section~\ref{sec:simplified}. 
Rather surprisingly, the hidden highlighted symmetries we will find
have spacetime signatures.

Let us first explain why we call these symmetries hidden. 
In Section~\ref{sec:gen:examples}, 
we solved the symmetry condition \eq{eq:gen:ex:gen_eq} for $N=3$.
There, the matrix $T$ was assumed to be real and antisymmetric, 
and the $SO(2)$ symmetry turned out to be the 
unique possibility.
This does not change, even if we consider $T$ to be an arbitrary real matrix, and we still 
get only the $SO(2)$ symmetry as the unique possibility. It will turn out that not the fundamental
dynamical variable of the theory, $P$, but rather the extended one (mentioned before in Section~\ref{sec:method}), $\tilde P$, 
will have a  spacetime-like symmetry.

Following the line of thought from the first paragraph of this section, let us consider $\tilde P$ instead of $P$, where $\tilde P$
is the real symmetric rank-3 tensor parametrizing the action \eq{eq:gen:ex:N=3_S_symmetric}
including the part with $\tilde \phi$:
\[
\begin{split}
&\tilde P_{113}=\tilde P_{223}=\frac{x}{3}, \\
&\tilde P_{333}=y, \\
&\tilde P_{114}=\tilde P_{224}=\tilde P_{334}=\frac{1}{3},\\
&\tilde P_{444}=-k,
\end{split}
\]
where $k=\frac{4}{27 \lambda}$, we have regarded $\tilde \phi=\phi_4$, 
and the other components up to permutations are zero.
We now want to find the solutions to the symmetry condition,
\[
T_{aa'}\tilde P_{a'bc}+T_{bb'}\tilde P_{ab'c}+T_{cc'}\tilde P_{abc'}=0
\label{eq:ttildep}
\] 
with arbitrary 4 by 4 real matrices $T$ (not restricted to be antisymmetric). 

The number of the entires of $T$ is 16, and it is easy to solve the condition by computers. 
For generic $x,y$, we again find only the $SO(2)$ matrix \eq{eq:so2mat} as the solution.
However, if we assume $y=x-k\, x^3$, 
namely the second equation in \eq{eq:gen:ex:N=3_critp_2} for the continuous critical points,
we obtain the following $T^{(2)}$ and $T^{(3)}$ in addition to the $SO(2)$ generator \eq{eq:so2mat}
as the solutions to \eq{eq:ttildep}:
\[
T^{(2)}_{13}=T^{(3)}_{23}=1,\ T^{(2)}_{14}=T^{(3)}_{24}=-x,\ T^{(2)}_{31}=T^{(3)}_{32}=-1+\frac{3}{2}k\,x^2,\ T^{(2)}_{41}=T^{(3)}_{42}=-\frac{3}{2}k\, x,
\]
where the other components are zero. One can check that they satisfy the following algebra, 
\[
\begin{split}
&[T^{(1)},T^{(2)}]=-T^{(3)},\\
&[T^{(1)},T^{(3)}]=T^{(2)},\\
&[T^{(2)},T^{(3)}]=(3 k\, x^2-1)\, T^{(1)},
\end{split}
\]
where $T^{(1)}$ denotes the $SO(2)$ generator \eq{eq:so2mat}.
As shown in Section~\ref{sec:gen:examples}, the strong peaks appear if $R^2=\frac{2x-3y}{x^3}> 0$, which 
implies nothing but the positivity of the coefficient $3 k\, x^2-1$ in the last line. 
Thus we have found an $SO(1,2)$ symmetry formed by $T^{(1,2,3)}$
as the highlighted symmetry on the strong peaks.

On the fixed points of the classical Hamiltonian vector flow the symmetry is enhanced even further. 
The flow is given by
${\pazocal H}_3$ \eq{eq:flowofh3},
which is drawn in Figure~\ref{fig:gen:ex:P_repr:N=3:SO(2)_contour_xy}.
There exist two kinds of fixed points, one at (i) $x=\pm 1/\sqrt{3k}$ and 
the other at (ii) $x=\pm 2/\sqrt{3k}$, on the curve $y=x-k\, x^3$.
In each case, in addition to the above $T^{(1,2,3)}$ with the substitution of the values of $x$,
there exists another symmetry transformation which solves \eq{eq:ttildep}: 
\[
\hbox{(i)}&:\ 
T^{\hbox{i}}_{11}=T^{\hbox{i}}_{22}=\frac{1}{2},\ T^{\hbox{i}}_{33}=1,\ T^{\hbox{i}}_{34}=\mp\frac{2}{\sqrt{3 k}},
\ T^{\hbox{i}}_{43}=\mp\sqrt{3k},\\
\hbox{(ii)}&:\ T^{\hbox{ii}}_{11}=T^{\hbox{ii}}_{22}=1,\ T^{\hbox{ii}}_{33}=-1,\ T^{\hbox{ii}}_{34}=\mp\frac{2}{\sqrt{3 k}},
\ T^{\hbox{ii}}_{43}=\mp\sqrt{3k},
\]
where the other components vanishes.
The algebras formed by them are given by
\[
(\hbox{i})&:\ [T^{(1)},T^{\hbox{i}}]=0,\ [T^{(2)},T^{\hbox{i}}]=\frac{3}{2}T^{(2)},\ [T^{(3)},T^{\hbox{i}}]=\frac{3}{2}T^{(3)},\\ 
(\hbox{ii})&:\ [T^{(1,2,3)},T^{\hbox{ii}}]=0. 
\]

For general $N$ with all $x_i=x$ in the simplified model discussed in 
Section~\ref{sec:simplified},
one can find the similar symmetry structures as above. 
The symmetry $SO(N-1)$ for general $x,y$ is 
enhanced to the hidden $SO(1,N-1)$ symmetry at the strong peaks on
$y=x-k\, x^3$, and there is an addition of another 
symmetry on each fixed point of the flow.

Presently we do not have a reliable physical interpretation of these highlighted hidden spacetime symmetries. 
These symmetries become apparent only after including the $\tilde \phi$ direction in the 
discussion of symmetries. 
This seems to suggest that this extra direction corresponds in some way to an implicit time-direction in the theory.
This interpretation seems justified as the extra terms with $\tilde \phi$ 
are originally introduced to satisfy the Hamiltonian constraint.
Moreover, the topology of the critical points in \eq{eq:gen:ex:N=3_critp_2} has the interesting structure
of a circle which changes its radius in the $\tilde \phi$ direction. 
Though this looks like a time evolving system, 
the integration variables $\phi$ and $\tphi$ have at this stage no physical interpretation. 
It is clear however that these
emergent spacetime symmetries will play some important roles in the spacetime interpretation of the dynamics
of the quantum CTM, which is yet to be explored.

\section{Asymptotic behaviour}\label{sec:asymptotic}
 The asymptotic behaviour of the wave function \eq{eq:review:Psi} is also important to analyze. A sufficiently fast damping wave function at infinity would mean that the wave function is normalizable, at least up to singularities in the finite $P$ regime.
 Furthermore, the wave function turns out to have some nontrivial asymptotic behaviour worth investigating. In this section we will first show an easy example of the asymptotic behaviour for a case with $N=3$ showing some of the nontrivial behaviours, then we will present a general scaling argument, which will be compared with some numerical results.
The scaling argument seems to explain the asymptotic behavior in most cases, 
 but will turn out not to cover all.
 In Appendix~\ref{app:normalizable}, we show in part the normalizablity of the 
 wave function at infinity, based on the analysis of the asymptotic behaviors discussed in this section.
 
 The example was introduced in Section~\ref{sec:gen:examples} and the resulting wave function is given in \eqref{eq:gen:ex:p1p32_exact}. Looking at the integrand, 
 the asymptotic behaviour with respect to $x,\ y$ and $z$ naively seems to be
 \begin{align}
    \begin{split}
     \Psi(x\rightarrow\infty) &\sim x^{-1},\\
     \Psi(y\rightarrow\infty) &\sim y^{-1/3},\\
     \Psi(z\rightarrow\infty) &\sim z^{-2/3}.
       \end{split}
       \label{eq:asymp:xy_1}
 \end{align}
 However, the last asymptotic behavior is not right, and it is actually given by
 \begin{equation}
     \Psi(z\rightarrow\infty) \sim z^{-1/2},\label{eq:asymp:z_1}
 \end{equation}
 which can easily be checked by the numerical method given in the preceding sections.
 The reason for the difference from the naive expectation from the integrand
 is that the integrand is not uniformly convergent due to the pole.
 Therefore, in general, one has to carefully look into the integral 
 to know the asymptotic behaviors in various infinite directions of the parameters. 
In the following section, we will give a scaling argument
which would be applicable to most cases, including the above
examples.
However, there do exist exceptional cases which cannot be 
understood simply by the scaling argument, and 
a more general method must be pursued in the future study.
 
 \subsection{Scaling argument}
 In this subsection, we will describe a scaling argument which 
 explains the asymptotic behaviors of the wave function for most cases of $P$.
 Let us consider the wave function \eq{eq:review:Psi}, 
 \begin{equation}
     \Psi(P) = \int_{\mathbb{R}^{N+1}} d\phi d\tphi \, e^{i(P \phi^3 + \phi^2 \tphi - k \tphi^3)},\label{eq:asymp:psi}
 \end{equation}
where $k=\frac{4}{27 \lambda}$ is assumed to be a positive constant.
Throughout this subsection, it is implicitly assumed that the integration is appropriately 
defined by the prescription described in Section~\ref{sec:method}.

A natural rescaling of the integration variables would be given by
\[
\phi_a \rightarrow |P|^{-\frac{1}{3}} \phi_a \ (a=1,2,\cdots,N),
\label{eq:naiverescale}
\]
where $|P|=\sqrt{P_{abc}P_{abc}}$. 
By the rescaling,  the action
in \eq{eq:asymp:psi} is transformed as 
\[
P\phi^3+\phi^2 \tilde \phi-k \tilde \phi^3 \rightarrow |P|^{-1} P\phi^3+|P|^{-\frac{2}{3}} \phi^2 \tilde \phi-k \tilde \phi^3.
\label{eq:actionwithP}
\]
If one naively assumes that the middle term can be neglected in the large-$P$ limit,
$\Psi$ will be estimated as 
\[
\Psi(P)\sim |P|^{-\frac{N}{3}} \int_{\mathbb{R}} d\tilde\phi\ e^{-ik\tilde \phi^3}
\int_{\mathbb{R}^N} d\phi \ e^{i|P|^{-1}P\phi^3}.  
\label{eq:productform}
\]
The first integral takes a finite non-zero value, and the last integral does not
depend on the overall scale of $P$. Therefore, when the last integral takes a finite non-zero value, 
the asymptotic behavior of $\Psi$, in which $|P|$ is taken infinitely large with constant 
$P_{abc}/|P|$, will be given by
\[
\Psi(P)\sim const.\  |P|^{-\frac{N}{3}},
\label{eq:asymPsi}
\]
where the overall factor is a function of $P_{abc}/|P|$.
The crucial assumption here is 
that the last integral of \eq{eq:productform} takes a finite non-zero value.
In most values of $P$, this will be true, and the asymptotic behavior will be given by \eq{eq:asymPsi}. 
However, there exist values of $P$ for which this is not true, and 
there actually exist rich varieties of asymptotic behaviors other than that.

A simple example with different asymptotic behaviors 
can be obtained from the calculable model discussed in Section~\ref{sec:simplified}. By setting $y=0$ 
in the model, we have
\[
P\phi^3=\sum_{i=1}^{N-1} x_i \phi_i^2 \phi_N.   
\]
If a uniform rescaling $\phi_a\rightarrow |x|^{-\frac{1}{3}} \phi_a$ with $|x|=\sqrt{\sum_{i} x_i^2}$ 
for all the $\phi_a\ (a=1,2,\cdots,N)$ is performed (as prescribed by \eq{eq:naiverescale}),
and the limit  $|x|\rightarrow \infty$ is taken, one obtains
\[
P\phi^3+\phi^2 \tilde \phi-k \tilde \phi^3 \rightarrow \sum_{i=1}^{N-1} \frac{x_i}{|x|} \phi_i^2 \phi_N-k \tilde \phi^3,
\]
where the middle term has naively been assumed to be negligible in the limit. 
The right-hand side 
contains $\phi_N$ in a linear form,  and this causes trouble. 
In fact, by performing the $\phi_N$ integration (and the $\tilde \phi$ integration), one obtains 
\[
\Psi\sim const.\  
|x|^{-\frac{N}{3}}\int_{\mathbb{R}^{N-1}} \prod_{i=1}^{N-1} d\phi_i \ \delta\left( \sum_{i=1}^{N-1} \frac{x_i}{|x|} \phi_i^2\right).
\]
Further integrations over $\phi_i\ (i=1,2,\cdots,N-1)$ 
will be in trouble: the integral vanishes or diverges, depending on the signs of $x_i$,
where the only finite case is $N=3$ with $x_{1,2}$ having the same sign.

A right way to deal with the above example is to consider a different rescaling than \eq{eq:naiverescale}. 
To get a finite convergent result of the integration, 
the action has to take an appropriate form in the limit.
In the present case, $\phi_N^2\tilde \phi$ term contained in the middle term would
be important in the limit, 
because it would prevent $\phi_N$ to appear only linearly, avoiding the trouble above.  
So, rather than uniformly rescaling all the $\phi_a$ like \eq{eq:naiverescale}, 
let us consider the following asymmetric rescaling, 
\[
\phi_i\rightarrow |x|^{-\frac{1}{2}} \phi_i\ (i=1,2,\ldots,N-1),\ \phi_N\rightarrow \phi_N.
\label{eq:exampleasymrescale}
\] 
By doing this rescaling and taking the limit $|x|\rightarrow \infty$, one obtains
\[
P\phi^3+\phi^2 \tilde \phi-k \tilde \phi^3 \rightarrow \sum_{i=1}^{N-1} \frac{x_i}{|x|} \phi_i^2 \phi_N
+\phi_N^2 \tilde \phi-k \tilde \phi^3,
\]
where $\phi_N^2 \tilde \phi$ indeed remains.
Then, the expression,
\[
\Psi\sim |x|^{-\frac{N-1}{2}} \int_{\mathbb{R}^{N+1}} d\phi d \tilde \phi\ \exp\left[i\left(
\sum_{i=1}^{N-1} \frac{x_i}{|x|} \phi_i^2 \phi_N+\phi_N^2 \tilde \phi-k \tilde \phi^3
\right)\right],
\label{eq:asymsolvable}
\]
gives the asymptotic behavior $\Psi\sim const.\ |x|^{-\frac{N-1}{2}}$,
if the integral takes a finite non-vanishing value.
This is actually the correct asymptotic behavior, if all the $x_i$ have the same sign.
In fact, the first line of \eq{eq:asymp:xy_1} corresponds to the $N=3$ case of 
\eq{eq:asymsolvable}.
However, if the signs of $x_i$ are mixed, the integral in \eq{eq:asymsolvable} 
is divergent: the integral cannot be defined as a strictly convergent integration, as explained in the following.
The $\phi_i$ integrations in \eq{eq:asymsolvable} are Gaussian and diverge for $\phi_N=0$.
Therefore, to define the integral properly, one needs to deform the integration contour of $\phi_N$ 
in the vicinity $\phi_N\sim 0$.
For the convergence of $\phi_i$ integration, the contour must be deformed as 
$\phi_N \rightarrow \phi_N + i \epsilon\, \hbox{Sign}(x_i)$
with a small positive $\epsilon$ in the vicinity $\phi_N\sim 0$.
However, this can be done consistently for all $\phi_i$, only if all $x_i$ have the same sign.

In the case of mixed signs of $x_i$, one must consider another rescaling,
\[
\phi_i\rightarrow \phi_i\ (i=1,2,\ldots,N-1),\ \phi_N\rightarrow \frac{1}{|x|}\phi_N.
\]
Then, one obtains
\[
\Psi\sim \frac{1}{|x|}  \int_{\mathbb{R}^{N+1}}  d\phi d \tilde \phi\ \exp\left[i\left(
\sum_{i=1}^{N-1}\left( \frac{x_i}{|x|}\phi_N +\tilde \phi \right) \phi_i^2-k \tilde \phi^3
\right)\right].
\label{eq:psiasymsec}
\]
In this case, the divergence of the $\phi_i$ integrations can be avoided by deforming the integration contour of 
$\phi_N$ in the manner mentioned above in the 
vicinity of $\frac{x_i}{|x|}\phi_N +\tilde \phi\sim 0$.
In the present case, unless $\tilde \phi=0$, the locations of $\phi_N$ are 
different for different $x_i$, and therefore
there are no contradictions like the former case. 
Moreover, the point $\tilde \phi=0$ can be circumvented by 
adding a small positive imaginary value to $\tilde \phi$ in the vicinity of $\tilde \phi \sim 0$ without ruining the 
convergence of the $\phi_i$ integrations. 
Here $x_i$ must have mixed signs for the expression \eq{eq:psiasymsec} to be 
useful, because otherwise the integral vanishes, which can be proven by
deforming the integration contour of $\phi_N$ to the positive imaginary infinity.
Thus the expression \eq{eq:psiasymsec} shows
that the asymptotic behavior is given by $const.\,/|x|$ for the mixed case.
This turns out to be the correct one for $N\geq 4$, while
$N=3$ is exceptional, and will be discussed at the end of this subsection.

Let us summarize our scaling argument in general terms. 
We want to obtain the asymptotic behavior of $\Psi$ for
$|P|\rightarrow \infty$ with all $P_{abc}/|P|$ being fixed. Let us consider a rescaling,
\[
\phi_a\rightarrow |P|^{-w_a} \phi_a\ (a=1,2,\cdots,N)
\label{eq:generalrescale}
\] 
with some weights $w_a$.
Then, $\Psi$ is transformed to
\[
\Psi=  |P|^{-\sum_{a=1}^{N} w_a} \int_{\mathbb{R}^{N+1}}  d\phi d\tilde \phi\ 
\exp \left[ i\left(
|P|^{-w_a-w_b-w_c}P_{abc} \phi_a \phi_b \phi_c + |P|^{-2 w_a} \phi_a \phi_a \tilde \phi -k \tilde \phi^3\right)
\right],
\label{eq:generalint}
\] 
where the repeated indices are assumed to be summed over. 
This leads to the following two conditions;
(i) The action in \eq{eq:generalint} has a finite limit in $|P|\rightarrow \infty$;
(ii) The integral takes a finite non-zero value for the limiting action.  
Under these conditions, one obtains the asymptotic behavior,
\[
\Psi\sim  const.\, |P|^{-\sum_{a=1}^{N} w_a}.
\label{eq:generalasymp}
\]

If one cannot find the set $\{w_a;a=1,2,\ldots, N\}$ which satisfies the two conditions above, 
the scaling argument cannot be applied. In fact, there exists such a counter example:
The $N=3$ case with $x_{1,2}$ having different signs.
To see this, let us perform the $\phi_i$ integrations in \eq{eq:psiasymsec}.
Then one obtains
\[
\Psi\sim \frac{const.}{|x|}\int d\phi_N d\tilde \phi\, \prod_{i=1}^{N-1} \frac{1}{\sqrt{
-i \left( \frac{x_i}{|x|} \phi_N + \tilde \phi \right)
}} \exp \left(-i k \tilde \phi^3 \right). \label{eq:app:norm:counter_example}
\]
For $N=3$, the integral over $\phi_N$ has a logarithmic divergence at infinity, and 
the condition (ii) is violated.
In fact, a numerical study shows that the asymptotic behavior of $\Psi$ is given 
by $(c_1+c_2 \log|x|)/|x|$ with constants $c_{1,2}$,
which has the form of the naive expression $1/|x|$ accompanied with a logarithmic correction.
Such a logarithmic correction cannot be treated by the present scaling argument. 
Therefore our scaling argument is not general enough to cover all the cases,
though it seems applicable to most cases. 

 \subsection{Examples}
 The general argument in the preceding subsection can now be used to explain the behavior of \eqref{eq:asymp:xy_1} and \eqref{eq:asymp:z_1}. First, for the large $x$-behavior we let $y$ and $z$ be fixed. This gives the leading term of the action as 
 \begin{equation*}
     x(\phi_1^2+\phi_2^2)\phi_3.
 \end{equation*}
 This corresponds to the first case of the example in the preceding subsection, namely
 $N=3$ with $x_{1,2}$ having the same sign. Therefore, one obtains the behavior $1/x$, as 
can be extracted from \eq{eq:asymsolvable}, in agreement with \eqref{eq:asymp:xy_1}.

 Let us next consider the $y\rightarrow \infty$ limit with $x,z$ being fixed. 
 The leading term is given by
 \begin{equation}
     y \phi_3^3.
 \end{equation}
For this case, one is lead to take $w_3=1/3$, because the condition (i) requires $w_3\geq 1/3$, but
$w_3 > 1/3$ violates the condition (ii). 
As for the other weights, one should take $w_{1,2}=0$, so that the limit of the action is
 given by $(\phi_1^2+\phi_2^2)\tilde \phi+ \phi_3^3 -4 \tphi^3/27 \lambda$. 
If one would take $w_{1,2}>0$, the integral of the limit would violate the condition (ii),
because the term $(\phi_1^2+\phi_2^2)\tilde \phi$ is needed for the convergence of the integral.
Then, one obtains $\Psi\sim y^{-1/3}$ in agreement with \eq{eq:asymp:xy_1}.

The last case is $z\rightarrow \infty$ with $x,y$ being fixed.
The leading term is
 \begin{equation}
     z \phi_1^2 \phi_3.
 \end{equation}
 In a similar way as above, one can determine $w_1=1/2,\ w_2=w_3=0$,
 which gives $\Psi\sim z^{-1/2}$ in agreement with \eqref{eq:asymp:z_1}. 
 
The asymptotic behavior for $N=3$ are summarized in Table~\ref{tab:asymp}.
In particular, there are cases, in which there exists the exceptional logarithmic
correction discussed at the end of the preceding subsection.
 \begin{table}
 \centering
  \begin{tabular}{ | c | c |}
    \hline
    Term & Asymptotic behavior\\
    \hline
    $(\phi_1^2+\phi_2^2)\phi_3$ & $z^{-1}$  \\
    \hline
    $\phi_3^3$ & $z^{-\frac{1}{3}}$  \\
    \hline
    $(\phi_1^2-\phi_2^2)\phi_3$ & $c_1 z^{-1}+c_2 z^{-1} \log(z) $ \\
    \hline
    $\phi_1^2\phi_3$ & $z^{-\frac{1}{2}}$  \\
    \hline
    $\phi_1\phi_2\phi_3$ & $c'_1 z^{-1}+c'_2 z^{-1} \log(z) $  \\
    \hline
  \end{tabular}
\caption{This table shows the asymptotic behavior of $\Psi$ for $N=3$
when the coefficient $z$ of the term shown in the first column is taken to
infinity. $c_{1,2},c'_{1,2}$ are some numerical constants.
The coincidence of the form between the third and the fifth rows can be understood by
an orthogonal transformation $\phi_1\rightarrow \frac{1}{\sqrt{2}}(\phi_1+\phi_2),\ 
\phi_2\rightarrow \frac{1}{\sqrt{2}}(\phi_1-\phi_2)$.}
\label{tab:asymp}
 \end{table}

 
Another simple example with non-trivial behavior is given by $P$ 
with a chain-like structure,
\[
P\phi^3=|P|\left(\phi_1 \phi_{2}^2+\phi_{2} \phi_{3}^2+\phi_{3} \phi_{4}^2+\cdots+
\phi_{N-1}\phi_N^2\right).
\label{eq:chain}
\]
Here, since $\phi_1$ is only linearly coupled, we should take $w_1=0$ 
to keep $\phi_1^2 \tilde \phi$ of the action in the limit. 
As for the other $\phi_a$, we should take
\[
w_2=\frac{1}{2},\ w_3=\frac{1}{4}, \ w_4=\frac{3}{8}, \cdots,
w_N=\frac{1}{3}\left(1+(-1)^N 2^{-N+1}\right),
\]
to cancel the overall factor $|P|$ of \eq{eq:chain}.
Then, the asymptotic behavior is determined to be
\[
\Psi\sim const. \, |P|^{-\sum_a w_a}=const.\,
|P|^{-\frac{1}{9}\left(3N-2+(-1)^N 2^{-(N-1)}\right)}.
\]
We have numerically checked the behaviors for some small $N$'s. 

As seen above, the wave function has rich asymptotic behaviors depending on the directions
of infinity, and it seems an interesting non-trivial question how to classify all the possibilities.

\section{Summary and future problems}
In this paper, we have studied in some detail 
the profile of a wave function 
which exactly solves all the quantum constraints of the canonical tensor model (CTM)  
for general $N$ \cite{Narain:2014cya}.
We have found the preference of symmetric configurations,
whose mechanism was described in the previous paper \cite{Obster:2017pdq}
for general settings. 
This preference has been found to occur only for $\lambda>0$, where $\lambda$ is a constant
in the ``Hamiltonian'' constraint of the CTM and is known to correspond to the cosmological constant 
for $N=1$.  
Surprisingly, we have found some symmetries with indefinite (spacetime-like) signatures associated
with the preferred configurations, not only the ones with positive definite (space-like) signatures. 
Since symmetries will determine the global characters of spacetimes,
the results are encouraging toward showing spacetime emergence in the CTM. 
We have also studied the asymptotic behaviors of the wave function for large values of $P$, and have found some rich structure.

An important technical detail in our work was to give the precise definition of the wave function, which had
rather formally been given in a previous paper~\cite{Narain:2014cya}.
The wave function has the expression of an integration over $N$ real variables, and 
the integrand oscillates infinitely fast at infinity with a constant modulus. 
In other words, the integral is a sort of multi-variable extension of the integral expression of the Airy function, 
and has conditionally convergent limits for generic values of the parameters contained in
the integrand. 
To properly handle this rather delicate integral, 
we introduced a regularization, the so-called $\epsilon$-prescription, 
and took its vanishing limit by properly deforming the integration contour. 
Then, the obtained expression of the wave function was analyzed mainly by numerical methods,
as well as partly by analytical methods for simplified settings.

We have found some rich structure of the peaks and the asymptotic behavior of the wave function,
which quickly become more and more complicated as $N$ becomes larger.  
Our present way of analysis, which is largely relying on numerical methods,
cannot provide thorough understanding of the properties of the wave function.    
Therefore there remain a long list of questions toward the full understanding of the quantum CTM.
 We considered only the particular wave function which has the most familiar Lorentzian form
and is valid for general $N$, but there exist other possibilities of the wave 
functions.
We would need to argue more strongly for the present particular choice of the wave function, 
or have to equally well consider the other possibilities in the CTM \cite{Narain:2014cya}
and the more fundamental model \cite{Narain:2016sqn}. 
We only considered the orthogonal group symmetries 
with the vector representations as highlighted symmetries,
but other representations and Lie-groups are also possible and interesting.  
For example, to describe an emergent two-sphere 
the expected highlighted configuration would take a form like
$P_{(l_1,m_1)(l_2,m_2)(l_3,m_3)}\sim \int d\omega\ Y_{l_1}^{m_1}(\omega)Y_{l_2}^{m_2}(\omega)
Y_{l_3}^{m_3}(\omega)$, where $Y_l^m(\omega)$ denotes the spherical harmonics,
and the irreducible representations labeled by $l$ run from spin-zero to a cut-off.
This interpretation is an area being left for later study.
The other Lie groups with real orthogonal representations, which can be embedded in 
the $O(N)$ matrices, are also interesting to be explored. 
The surprising appearance of spacetime signatures associated to the preferred configurations
should obviously be understood more deeply. 
We only studied the profile of the wave function, but  
we rather have to perform integrations over $P$ to evaluate physical quantities like  
$\int dP \, |\Psi(P)|^2 {\cal O}(P)$ with an observable ${\cal O}(P)$. 
Without doing this, we do not even determine whether the highlighted peaks are 
really physically sensible or not. This is also important to see whether the divergences at the 
peaks are physically harmless or not.
It seems necessary to develop more effective and systematic methods to answer these questions.

\vspace{1cm}
\section*{Acknowledgements}
The work of N.S. is supported in part by JSPS KAKENHI Grant Number 15K05050. 
The work of D.O. is supported in part by the Hendrik Mullerfonds.
We would like to thank some of the participants for stimulating discussions
at the workshop ``Discrete Approaches to the Dynamics of Fields and Space-Time'' held in APCTP, where the contents of this paper were presented.  
 
 \appendix

\section{The derivation of the wave function}
\label{app:wavefunction}
In this section of the appendix, we will show the derivation of the wave function \eq{eq:review:Psi} \cite{Narain:2014cya} to make this paper self-contained.
In the derivation, the validity of partial integrations is essentially important. 
This can be assured by taking appropriate integration contours or the appropriate prescription of 
regularization as taken in Section~\ref{sec:method}.  
The derivation of the other wave function \eq{eq:exactwithnolambda} is similar.
We will also describe the result of the consistency checks of the numerical evaluation of  
the wave function \eq{eq:gen:regwavefunction}, 
which is mentioned in the end of Section~\ref{sec:method}.

From \eq{eq:quantumHamiltonian} and \eq{eq:review:CTMwheelerdewitt},
the Hamiltonian constraint equations in the $P$ representation are given by
\[
(P_{abc}P_{bde} D^P_{cde}+\lambda_H P_{abb} - \lambda D^P_{abb})\Psi_{phys}(P)=0,
\label{eq:wdwcosmo}
\]
where $D^P_{abc}$ are the derivative operators with respect to $P_{abc}$ with 
the following normalization,
\[
D^P_{abc} P_{def}= \sum_{\sigma} \delta_{a\sigma_d}\delta_{b\sigma_e}\delta_{c\sigma_f}, 
\]
where $\sigma$ denote the permutations of $d,e$ and $f$.
To get a solution, let us consider an ansatz,
\[
\Psi(P)=\int_{\pazocal C} d\phi \, f(\phi^2) e^{i P\phi^3},
\label{eq:appD:ansatz}
\]
which was motivated from the close connection between the CTM and the randomly connected tensor networks 
\cite{Sasakura:2015xxa,Sasakura:2014yoa,Sasakura:2014zwa}.
Here, $f$ denotes a function to be determined below.
Applying the first operator in \eq{eq:wdwcosmo}, we obtain 
\[
P_{abc} P_{bde} D^P_{cde} \Psi(P)
&=6 i \int_{\pazocal C} d\phi \ P_{abc}P_{bde}\phi_c\phi_d\phi_e f(\phi^2) e^{i P\phi^3} \CR
&=2 \int_{\pazocal C} d\phi \ P_{abc} \phi_c f(\phi^2) \partial_b e^{i P\phi^3} \CR
&=-2 \int_{\pazocal C}  d\phi\ \partial_b\left(P_{abc} \phi_c f(\phi^2)\right) e^{i P\phi^3} \CR
&=-2 \int_{\pazocal C} d\phi \ \left( P_{abb} f(\phi^2) +2 P_{abc}\phi_b \phi_c f'(\phi^2)\right) e^{i P\phi^3},
\label{eq:applyPPDQ}
\]
where  $\partial_a$ denotes the derivative with respect to $\phi_a$, 
$f'$ the derivative of $f$ with respect to the argument, and 
we have performed some partial integrations with no boundary contributions, 
which are assumed to be valid
by appropriately taking ${\pazocal C}$.
The last term of \eq{eq:applyPPDQ} can further be computed as
\[
\begin{split}
\int_{\pazocal C} d\phi \  P_{abc}\phi_b \phi_c f'(\phi^2) e^{i P\phi^3}
&=\frac{1}{3 i}\int_{\pazocal C} d\phi \  f'(\phi^2) \partial_a e^{i P\phi^3} \\
&=\frac{2 i}{3}\int_{\pazocal C} d\phi \  \phi_a f''(\phi^2) e^{i P\phi^3}.
\end{split}
\label{eq:compthird}
\]
Now let us assume
\[
f''(x)=A x f(x)
\label{eq:condfx}
\]
with a numerical constant $A$. 
Then, the last expression in \eq{eq:compthird} can further be computed as
\[
\begin{split}
\int_{\pazocal C} d\phi \  \phi_a f''(\phi^2) e^{i P\phi^3}
&=A \int_{\pazocal C}d\phi \ \phi_a \phi^2 f(\phi^2) e^{i P \phi^3} \\
&=\frac{A}{6 i } D^P_{abb} \int_{\pazocal C} d\phi \ f(\phi^2) e^{i P \phi^3}.
\end{split}
\]
Finally, by collecting the expressions above, we obtain an identity satisfied by $\Psi(P)$,
\[
\left[ P_{abc} P_{bde} D^P_{cde} +2 P_{abb}
+\frac{4A}{9} D^P_{abb}\right] \Psi(P)=0.
\label{eq:WDWwithcos}
\]
This identity implies that a solution to \eq{eq:wdwcosmo} is given by
\eq{eq:relpsiQP},
if we put
\[
A=-\frac{9}{4} \lambda.
\]

As for \eq{eq:condfx}, the solution is given by the Airy function,
and an integral expression of $f$ can be given by
\[
f(x)=\int_{\tilde {\pazocal C}} d\tilde \phi\ \exp \left[i\left( x \tilde \phi +\frac{\tilde \phi^3}{3 A }\right) \right]
\label{eq:fairy}
\]
with an appropriate integration contour $\tilde {\pazocal C}$.
By putting this expression into the ansatz \eq{eq:appD:ansatz}, we obtain  \eq{eq:review:Psi}.

Let us finally check the momentum constraints. Similarly, by performing some partial integrations, 
we obtain
\[
\begin{split}
\hat {\pazocal J}_{ab}\Psi(P)
&=(P_{acd}D^P_{bcd}-P_{bcd}D^P_{acd}) \int_{\pazocal C} d\phi \ f(\phi^2) e^{i P\phi^3} \\
&=6 i  \int_{\pazocal C} d\phi \ f(\phi^2)(P_{acd} \phi_b\phi_c\phi_d- P_{bcd}\phi_a\phi_c\phi_d)
 e^{i P\phi^3} \\ 
&=2 i  \int_{\pazocal C} d\phi \ f(\phi^2)(\phi_b\partial_a- \phi_a\partial_b)
 e^{i P\phi^3}\\
&= -2 i  \int_{\pazocal C} d\phi \ \left(\partial_a (f(\phi^2)\phi_b)-  \partial_b (f(\phi^2)\phi_a)\right)
 e^{i  P\phi^3}\\
&=0.
\end{split}
\]
This proves $\Psi_{phys}$ satisfies the momentum constraints.

The integration region of the wave function considered in the 
text is a real plane, and hence the integrand does not damp at infinity. 
Nonetheless, the integration converges conditionally for generic $P$,
because the integrand oscillates infinitely fast at the infinity
of the integration region. 
As in Section~\ref{sec:general}, we treat this delicate integration by introducing 
the so-called $\epsilon$-prescription, which is often used in physics to regularize, 
and finally take the $\epsilon\rightarrow +0$ limit.
In this case, the above proof cannot be applied to the wave 
function, 
since there exists an extra contribution to the identity \eq{eq:WDWwithcos} 
from the regularization term. In general, it is not a trivial
question whether the extra contribution vanishes in the vanishing limit
of the regularization,
but the present case is easy to answer as follows.
If we add the regularization term, 
$-\epsilon \phi^2 -\epsilon \tilde \phi^2 $, to the exponent of \eq{eq:appD:ansatz},
the additional term to \eq{eq:WDWwithcos} turns out to be 
\[
\epsilon \int_{{\mathbb R}^{N+1}} d\phi d\tilde \phi\ \left( c_1 P_{abc}\phi_b\phi_c+c_2 \phi_a \tilde \phi\right)
e^{i\left( P\phi^3 +\phi^2 \tilde \phi-\frac{4}{27 \lambda} \tilde \phi^3\right) 
-\epsilon \phi^2 -\epsilon\tilde \phi^2 } 
\label{eq:correctionham}
\]
with numerical constants $c_1,c_2$. 
This vanishes in the $\epsilon \rightarrow 0^+$ limit for generic $P$,
because, irrespective of the additional function of $\phi,\tilde \phi$ in the integrand, 
the integration is still conditionally convergent due to the infinitely fast oscillations of the 
integrand at infinity, and the overall factor $\epsilon$ will make the breaking term to vanish.

\begin{figure}
\begin{center}
\includegraphics[scale=.3]{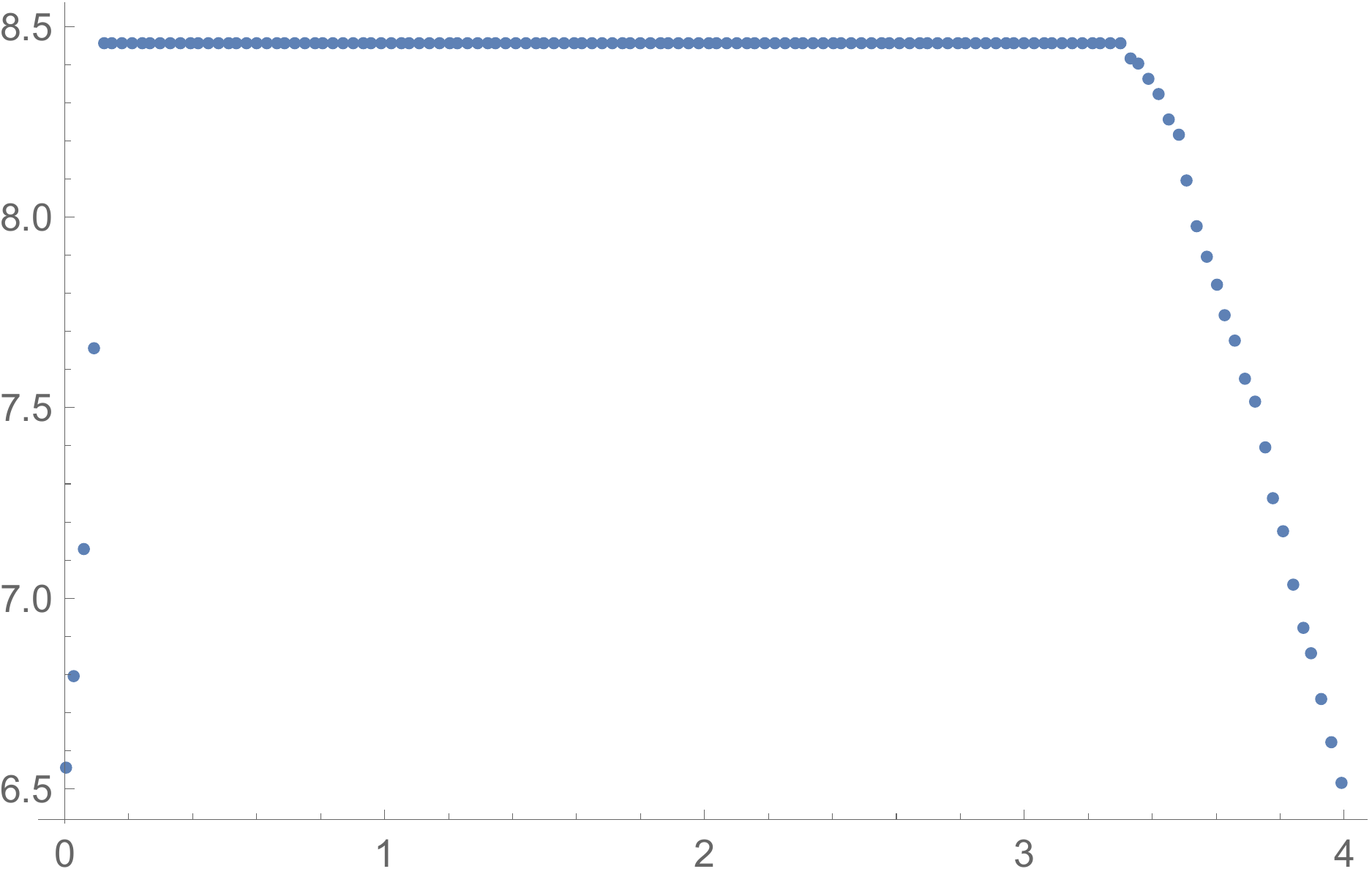}
\hfil
\includegraphics[scale=.3]{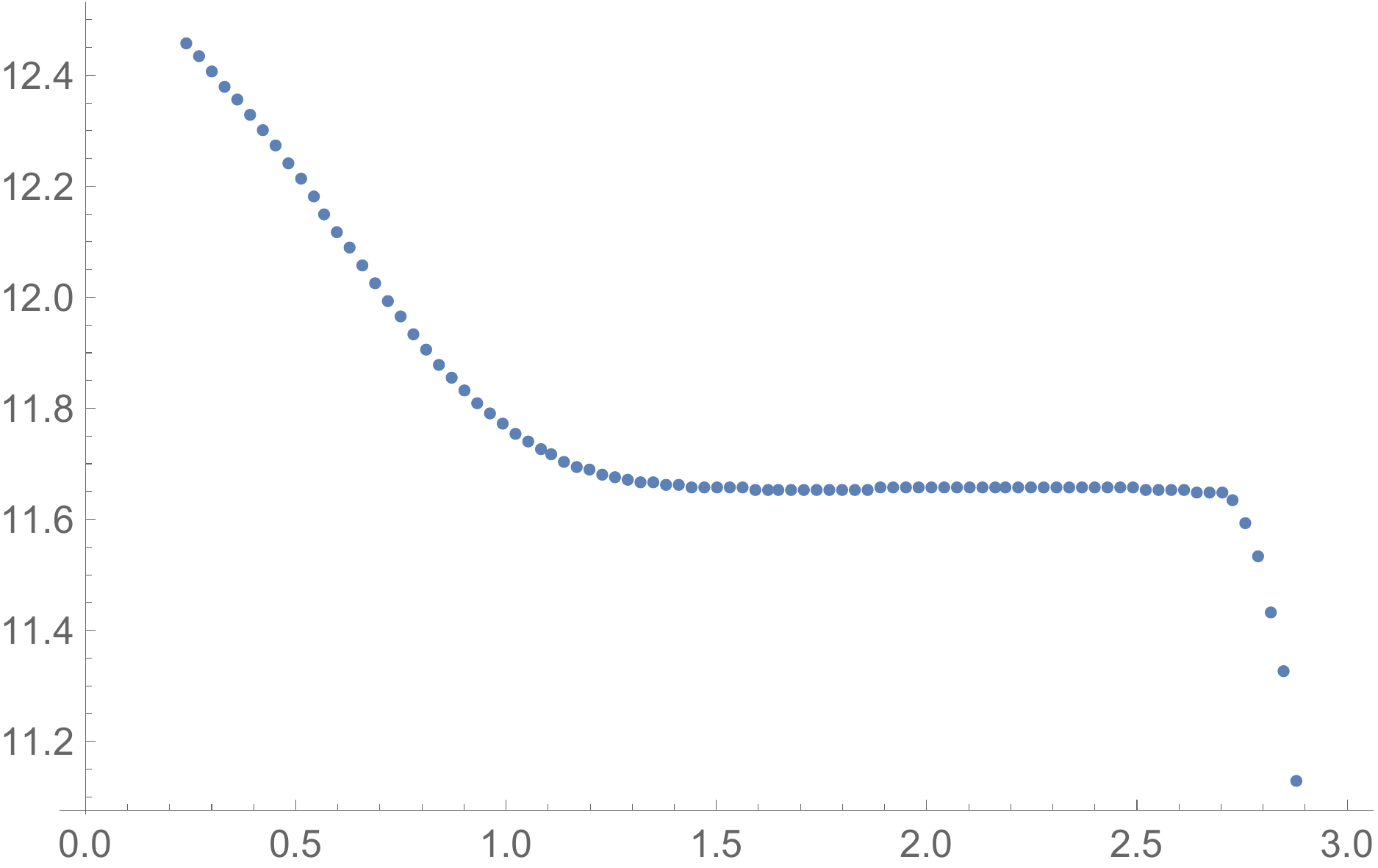}
\caption{The dependence of the numerical values of 
the wave function \eq{eq:gen:regwavefunction} on $\Delta$.
The left is for $N=2$ and $\lambda=1$ at a ``generic'' configuration $(P_{111},P_{112},P_{122},P_{222})=(1,1.1,1.2,1.3)$,
and the right for $N=3$ and $\lambda=1$ at $(P_{111},P_{112},\cdots,P_{333})=(1.0111,1.0222,1.0333,\cdots,1.111)$. 
The horizontal axis represents $-\log_{10}(\Delta)$, 
and the vertical axis the wave function. There exist substantial regions of constant values,
which are supposed to be the real values of the wave function.}
\label{fig:deltadep}
\end{center} 
\end{figure}

The above argument supports the validity of the $\epsilon$-prescription taken in
the text. As explained in Section~~\ref{sec:method}, the $\epsilon\rightarrow +0$ limit requires a deformation of the integration contour which should not change the values of the wave function 
due to the Cauchy theorem, if the deformation parameter $\Delta$ is small enough.   
To be sure, we performed some direct numerical checks of our method 
of computing \eq{eq:gen:regwavefunction}.
We studied the dependence of the wave function on $\Delta$
for $N=1,2,3$. Two of the results are shown in Figure~\ref{fig:deltadep}.
As shown, the graphs contain the substantial regions of constancy, the values of which
can be regarded as the real values of the wave function in the vanishing limit of the regularization. 
The deviations in the left region (larger $\Delta$) should come from 
that the deformed contours cross some branch cuts or singularities. 
The deviations in the right region (smaller $\Delta$) 
 come from the fact that the deformed contour is so close to the singularities 
that the numerical integrations suffer from large errors.

Another check was to see whether the key identity \eq{eq:WDWwithcos} is satisfied by 
the numerically computed wave function. We have obtained some satisfactorily small numbers of 
the violations for $N=1,2,3$. 
For example, we obtained the violation of order $\sim 10^{-8}$ for $N=2$
and $\Delta=0.1$, and $~\sim 10^{-2}$ for $N=3$ and $\Delta=10^{-1.8}$,
respectively, for the same parameters used in Figure~\ref{fig:deltadep}.
Though it is hard to judge whether these numbers can be regarded as zero, we also observed the tendency that
the violations became smaller when the optional parameters of the numerical integration command 
were chosen to produce more precise numerical values.

In the checks above, several hours of computation times were needed even for $N=3$ to get the
satisfactorily precise results shown in Figure~\ref{fig:deltadep}.\footnote{We performed the numerical computation by Mathematica 11. The
optional parameters of the numerical integration command had to be tuned to get better results. 
In our computation, it was important to take the option value 
``MaxErrorIncreases'' to be at least several ten thousands, 
while its default value is two thousands.}  
However, the slow speed to get precise values  
is not so problematic for our main purpose. 
This is because we are not interested in the values themselves, but in the qualitative behavior of the wave function
to see the highlighting phenomenon of symmetric configurations.
Therefore, in most cases, we can set the optional parameters 
in favor of the speed, sacrificing unnecessary preciseness.

\section{The difficulty of the $\lambda=0$ wave functions}
\label{appendix:sec:wavewithoutl}
In this appendix, we will explain the difficulty of the wave functions for $\lambda=0$,
namely \eq{eq:review:Psi} with $\lambda=0$ and \eq{eq:exactwithnolambda}. 

Let us first consider \eq{eq:review:Psi}.
By a change of variable $\tilde \phi \rightarrow |\lambda|^{1/3} \tilde \phi$ and taking 
the limit $\lambda\rightarrow 0$, one can see that $\phi$ and $\tilde \phi$ are decoupled. 
$P$ is coupled only with $\phi$, and the wave function is given by
\[
\Psi(P)=\int_{\mathbb{R}^N} d\phi\ e^{i P\phi^3}.
\]
By considering a rescaling of variables $\phi\rightarrow |P|^{-1/3} \phi$, it is obvious 
that the wave function splits into the radial and angular parts as
\[
\Psi(P)= |P|^{-\frac{N}{3}} \Psi(P_\Omega),
\]
where we have introduced a polar coordinate with $P_{\Omega\,abc}=P_{abc}/|P|$.  
Then, from \eq{eq:valoflambdah} and \eq{eq:relpsiQP}, the physical wave function is given by
\[
\Psi_{phys}(P)=|P|^{-\frac{N(N+2)(N+3)}{12}} \Psi(P_\Omega)^{\frac{(N+2)(N+3)}{4}}.
\]
Therefore, the wave function has a strong peak at the origin $P\sim 0$.
To be more precise, let us include the volume factor as well:
\[
\int dP\ |\Psi_{phys}(P)|^2
=\int \frac{d|P|}{|P|} dP_{\Omega}\ |P|^{-\frac{N(N+2)}{3}} |\Psi(P_\Omega)|^\frac{(N+2)(N+3)}{2},
\]
where we have used the fact that the dimension of the space of $P$ is given by $N(N+1)(N+2)/6$.
Therefore, $P=0$ is the most favorable configuration, which would have the physical meaning 
that there are no spaces.

This preference to $P=0$ is qualitatively understandable, 
because, at $P=0$, the integrand does not depend on $\phi$ and the integration trivially diverges. 
Not only to this global preference, but similar reasoning can also be applied to the partial cases that 
the configurations with $P_{abc}=0,\ \exists a, \forall b,c$
are relatively preferred, because then the integrand does not depend on $\phi_a$,
and the integration over $\phi_a$ diverges.  
Therefore, generally speaking, the configurations with less effective $N$ are preferred. 
This tendency of collapse   
will make the case of $\lambda=0$ difficult to be physically sensible.

Let us turn to \eq{eq:exactwithnolambda}. In a similar manner, we obtain
\[
\Psi(Q)=|Q|^{-\frac{N+2 \alpha}{3}} \Psi(Q_\Omega),
\]
where $\alpha=(N+3)(N-2)/8$. Then, 
\[
\int dQ\ |\Psi(Q)|^2=\int \frac{d|Q|}{|Q|} dQ_{\Omega}\ |Q|^{\frac{N^3+2 N^2-3 N+6}{6}} |\Psi(Q_\Omega)|^2.
\]
Since the exponent is positive for $N\geq 1$, the wave function diverges at infinity. This problem of this case seems be a ``conjugate dual'' to the former one; the wave function spreads out to infinity. 
It would be difficult to physically make sense of this case as well.

\section{The simple expression of the calculable model}
\label{app:reduce}
The integral we want to consider here is 
\[
\Psi=\int_{\mathbb{R}^{N+1}} \prod_{j=1}^Nd\phi_jd\tilde \phi
\exp \left(i S(\phi,\tilde \phi, x,y)-\epsilon (\phi^2+\tilde \phi^2)\right), 
\]
where $\epsilon >0$, $\phi^2=\sum_{i=1}^{N} \phi_i^2$, and 
\[
S(\phi,\tilde \phi,x,y)=\sum_{i=1}^{N-1} x_i \phi_i^2 \phi_N+y \phi_N^3+\phi^2 \tilde \phi - k \tilde \phi^3 
\label{eq:action}
\]
with $k=\frac{4}{27 \lambda}$. 

There are two difficult things in the actual evaluation of this integration. 
One is the multivariable integration, and the other is 
the $\epsilon\rightarrow +0$ limit. We will reduce the former to a single integration, and 
safely take the latter by deforming the integration contour of the remaining
single integration.

By doing the Gaussian integrations over $\phi_i\ (i=1,2,\ldots,N-1)$, one obtains 
\[
\Psi=\pi^\frac{N-1}{2}\int_{\mathbb{R}^2} d\phi d\tilde \phi \prod_{j=1}^{N-1} \frac{1}{\sqrt{\epsilon - i (x_j \phi+\tilde \phi)}}
\exp \left( i \left( \phi^2 \tilde \phi+y \phi^3 - k \tilde \phi^3\right)-\epsilon (\phi^2+\tilde \phi^2)\right),
\label{eq:integral}
\]
where $\phi_N$ has been replaced by $\phi$ for a notational simplification.

By dividing the integration region of $\phi$ into the positive and negative regions, the wave function \eq{eq:integral} 
can be expressed as
\[
\Psi=\pi^\frac{N-1}{2}\left(A_++A_-\right),
\]
where
\[
A_+&=\int_0^\infty d\phi \int_{-\infty}^\infty d\tilde \phi 
\prod_{j=1}^{N-1} \frac{1}{\sqrt{\epsilon - i (x_j \phi+\tilde \phi)}}
\exp \left( i \left( \phi^2 \tilde \phi+y \phi^3 - k \tilde \phi^3\right)-\epsilon (\phi^2+\tilde \phi^2)\right),
\label{eq:repofaplus} \\
A_-&=\int_{-\infty}^0 d\phi \int_{-\infty}^\infty d\tilde \phi 
\prod_{j=1}^{N-1} \frac{1}{\sqrt{\epsilon - i (x_j \phi+\tilde \phi)}}
\exp \left( i \left( \phi^2 \tilde \phi+y \phi^3 - k \tilde \phi^3\right)-\epsilon (\phi^2+\tilde \phi^2)\right).
\]
One can easily show that $A_+^*=A_-$ by performing the change of the variables, 
$\phi\rightarrow -\phi, \tilde \phi\rightarrow
-\tilde \phi$. Therefore, 
\[
\Psi=2 \pi^\frac{N-1}{2} \hbox{Re}\left[ A_+\right].
\label{eq:psiap}
\]

To compute $A_+$, let us perform the change of variable, $\tilde \phi\rightarrow \phi \tilde \phi$.
Then, we obtain
\[
A_+=\int_0^\infty d\phi \int_{-\infty}^\infty d\tilde \phi
\prod_{j=1}^{N-1} \frac{1}{\sqrt{\epsilon - i (x_j +\tilde \phi)}}\phi^{1-\frac{N-1}{2}}
\exp \left(-\left(\epsilon-  i \left( \tilde \phi+y  - k \tilde \phi^3\right)\right)\phi^3 \right),
\label{eq:intphi}
\]
where we have used the positivity of $\phi$. Here, the regularization 
parameter $\epsilon$ has been replaced,
keeping the same roles as in \eq{eq:repofaplus},
namely,  a suppression term at infinity and the choice of the branches of the square roots.
However, the replacement introduces a new singularity at $\phi=0$, as can be seen 
in the power of $\phi$ in the integrand of \eq{eq:intphi}.

As mentioned above, the integration \eq{eq:intphi} has a divergence 
at the endpoint $\phi=0$ for positive $N$. This singularity did not exist in the original 
integration before the replacement of the regularization parameter, and therefore 
has to be regulated in another way.
One way is to consider an analytic continuation of $N$ from negative to positive values.
 Then, by formally carrying out the $\phi$ integration, one obtains
\[
A_+=\frac{1}{3} \Gamma\left( \frac{5-N}{6}\right) \int_{-\infty}^\infty d \tilde \phi
\prod_{j=1}^{N-1} \frac{1}{\sqrt{\epsilon - i (x_j+\tilde \phi)}} \left(\epsilon-i h(\tilde \phi)\right)^\frac{N-5}{6},
\label{eq:apwitheps}
\]
where 
\[
h(\tilde \phi)=\tilde \phi+y  - k \tilde \phi^3.
\]
Here, the remaining $\tilde \phi$ integration is convergent, because the integrand behaves like $\sim \tilde \phi^{-2}$ as it tends to infinity. 

In \eq{eq:apwitheps}, the regularization parameter $\epsilon$ determines how to 
take the integration contour in relation with the branch cuts of the integrand.
By using the Cauchy theorem, one can take the $\epsilon\rightarrow 0^+$ limit by continuously deforming 
the integration contour $\pazocal C$ away from the real plane.
Here, the fractional powers of the integrand in \eq{eq:apwitheps} are supposed to be taken in the main branches.
Hence, the branch cuts associated to $\tilde \phi=-x_i$ extend in the direction of the negative 
pure imaginary as in Fig.\ref{fig:contourtilde}. 
If we first consider the case that $(N-5)/6$ is fractional,
there also exist branch cuts associated to the solutions to $h(\tilde \phi)=0$.
The relevant solutions are those on the real axis, and the branch cuts extend from there to
the negative or positive imaginary regions, depending on whether the signs of $h'(\tilde \phi)$ at the solutions are positive or negative, respectively.
The contour ${\pazocal C}$ should be taken so as to circumvent those branch cuts.
An example is shown in Fig.~\ref{fig:contourtilde}. 
With this understanding of the integration contour ${\pazocal C}$, one obtains an $\epsilon$-free expression,
\[
A_+=\frac{1}{3} \Gamma\left( \frac{5-N}{6}\right) \int_{\pazocal C} d \tilde \phi
\prod_{j=1}^{N-1} \frac{1}{\sqrt{ - i (x_j+\tilde \phi)}} \left(-i h(\tilde \phi)\right)^\frac{N-5}{6}.
\label{eq:ap}
\]

When $n=(N-5)/6$ is a non-negative integer, \eq{eq:ap} cannot be used,
because the allover factor (the gamma function) is divergent. 
Moreover, in this case,
the integrand has no singularities in the positive imaginary region, and 
one can deform  ${\pazocal C}$ to infinity to show that the integration 
vanishes because of the fast damping behavior $\tilde \phi^{-2}$ of the integrand.
Thus, the expression \eq{eq:ap} is actually indeterministic, $\Psi=\infty\cdot 0$.

To resolve this issue, let us take the analytic continuation in $N$ more carefully. 
Let us consider a perturbation of $N$ as
$N=6n+5+6 \alpha$ with an infinitesimal $\alpha$. The relevant formulas are
\[
\begin{split}
\Gamma\left( \frac{5-N}{6}\right)&=\frac{(-1)^{n+1}}{n!\, \alpha}+\cdots, \\
(\epsilon-i h)^{n+\alpha}&=(\epsilon-i h)^n e^{\alpha \ln(\epsilon-i h)}=(\epsilon-i h)^n (1+\alpha \ln(\epsilon-i h)+\cdots).
\end{split}
\]
Putting these into \eq{eq:ap} and taking the zeroth order in $\alpha$ (the lowest order $\alpha^{-1}$
vanishes because of the vanishing of the integration explained above), one obtains
\[
A_+=\frac{(-1)^{n+1}}{3 n!} \int_{\pazocal C}d\tilde \phi \prod_{j=1}^{N-1} \frac{1}{\sqrt{ - i (x_j+\tilde \phi)}} \left(-i h(\tilde \phi)\right)^n \ln
\left(-i h(\tilde \phi)\right),
\label{eq:apint}
\]
where the integration contour is taken in the same manner as previously.

The formula \eq{eq:psiap} with \eq{eq:ap} and \eq{eq:apint} was numerically compared with
the computation based on the generally applicable method (but rather slow due to the multivariable integration) 
explained in Section~\ref{sec:method}.  
We have found perfect agreement in all the cases we checked up to $N=5$,
supporting the validity of the derivation of the formula.

\section{Normalizability of the wave function in the large-$P$ region}
\label{app:normalizable}
In this appendix, we will discuss the normalizablity of the wave function
in the large-$P$ region, when the wave function has the asymptotic behavior which 
can be derived from the scaling argument discussed in Section~\ref{sec:asymptotic}.
This appendix does not fully prove the normalizability of 
the wave function in the large-$P$ region, because we assume the scaling argument, which
we know does not cover all the cases, as shown in Section~\ref{sec:asymptotic}. 
Therefore, this appendix proves the normalizability only in a part of the large-$P$ region,
which we expect should cover most of the configuration space.

As discussed in Section~\ref{sec:asymptotic}, the scaling argument assumes 
the existence of a scaling of $\phi_a$ which satisfies the conditions (i) and (ii).
One can easily derive the following set of necessary conditions for (i) and (ii):
\[
\begin{split}
&w_a+w_b+w_c\geq 1 \ \hbox{if }P_{abc}\neq 0,\  \forall\, a,b,c,\\
&w_a+w_b+w_c= 1, \ \exists\, a,b,c,\\
&w_a \geq 0, \ \forall\, a.
\label{eq:minimalcond}
\end{split}
\]
The first one is necessary for the condition (i), because, after the rescaling of $\phi$,
the term $P_{abc}|P|^{-w_a-w_b-w_c} \phi_a \phi_b \phi_c$ in the action should not diverge
in the asymptotic limit $|P|\rightarrow \infty$.
The third one is necessary with the same reason for $|P|^{-2w_a} \phi_a^2\tilde \phi$ 
in the action.
The second one is necessary for the condition (ii),
because at least one triple term must remain in the action in the asymptotic limit
for the convergence of the integral.
Here, note that it is not possible to keep all the $\phi_a^2 \tilde \phi$ terms in the action 
in the asymptotic limit for assuring the convergence of the integral,
because the first one of \eq{eq:minimalcond} requires at least one of $w_a$ must be positive.

As shown in Appendix~\ref{app:wavefunction},
the physical wave function is given by \eq{eq:relpsiQP}
with \eq{eq:valoflambdah}:
\[
\Psi_{phys}(P)=\Psi(P)^{\frac{1}{4}(N+2)(N+3)},
\label{eq:wavehermite}
\]
where $\Psi$ is the wave function \eq{eq:review:Psi}.
Therefore, if we assume the scaling argument, 
the asymptotic behavior of $\Psi_{phys}$ is given by
\[
\Psi_{phys} \sim const. \,|P|^{-\frac{1}{4}(N+2)(N+3) \sum_{a=1}^N w_a},
\label{eq:asymppsih}
\]
where we have used \eq{eq:generalasymp}.

Now, let us consider perturbations $\delta P_{abc}$ of $P_{abc}$, and 
qualitatively estimate the allowed range of the perturbations  
under the requirement that the asymptotic behavior keeps the same form \eq{eq:asymppsih}
with a given set $\{w_a;\ a=1,2,\ldots,N\}$.
First of all, if the perturbations $\delta P_{abc}$ of $P_{abc}$
satisfy $\delta P_{abc} |P|^{-(w_a+w_b+w_c)} \rightarrow 0$ in $|P|\rightarrow \infty$,
the limiting action after the rescaling does not change. 
Therefore, in this case, the wave function keeps the same asymptotic form as \eq{eq:asymppsih}
including the overall constant.
We can consider more general perturbations.
Because of (ii), the limiting action can be perturbed by certain finite amounts without
losing the convergence of the integral. Such perturbations are in the order of 
$\delta P_{abc} \sim |P|^{w_a+w_b+w_c}$.
In this case, the overall constant in \eq{eq:asymppsih} can change,
because it is determined by the integration value with the limiting action,
while the scaling behavior in $|P|$ keeps the same form.  
Thus we obtain the following qualitative estimation of the range of perturbations
which are allowed for the asymptotic behavior determined by a given 
set $\{w_a;\ a=1,2,\ldots,N\}$: 
\[
\delta P_{abc}\lesssim |P|^{w_a+w_b+w_c}. 
\label{eq:rangeofPinM}
\]  
Then, the contribution to the norm of $\Psi_{phys}$ from such a region, 
denoted below by $P_{\{w\}}$, can be estimated as
\[
\begin{split}
||\Psi_{phys}||^2_{P_{\{w\}}}&= \int_{P_{\{w\}}} \prod_{a,b,c=1 \atop a\leq b\leq c}^N dP_{abc}\  |\Psi_{phys}|^2 \\
&\sim \int \frac{d|P|}{|P|} \left(\prod_{a,b,c=1 \atop a\leq b\leq c}^N |P|^{w_a+w_b+w_c}\right) 
|P|^{-\frac{1}{2}(N+2)(N+3) \sum_{a=1}^N w_a} \\
&= \int \frac{d|P|}{|P|} |P|^{-(N+2)\sum_{a=1}^N w_a} \\
&\leq  \int \frac{d|P|}{|P|} |P|^{-\frac{1}{3}(N+2)} <\infty.
\end{split}
\label{eq:intinM}
\]
Here, from the first to the second line, we have used \eq{eq:asymppsih} and the range \eq{eq:rangeofPinM};
from the second to the third line, we have used the fact that each $w_a$ 
appears $\frac{1}{2}(N+1)(N+2)$ times
in the product; from the third to the last line, we have used $\sum_{a=1}^N w_a\geq \frac{1}{3}$,
which can be proved from \eq{eq:minimalcond}. 
The estimate \eq{eq:intinM} shows that $\Psi_{phys}$ is normalizable in the large-$P$ region of 
$P_{\{w\}}$. 

Note that the estimation above does not prove the normalizability of 
the wave function in the large-$P$ region. 
The obstacle is that we do not know exactly to what extent the $P_{\{w\}}$ 
cover the whole large-$P$ region, since we know that our 
scaling argument does not cover all the possible asymptotic behaviors,
as shown in Section~\ref{sec:asymptotic}. 
What we have shown in this appendix is merely that the normalizability of the wave function
at the large-$P$ region is assured at least in the vicinities of $P$ with the asymptotic 
behaviors consistent with the scaling argument. 
On the other hand, the result of this appendix seem to narrow down the possibilities of the 
breakdown of the normalizability at the large-$P$ region to the following two kinds of locations:  
The boundaries between different asymptotic regions, and  
the vicinities of the exceptional cases to our scaling argument.
Though the discussions in this appendix are qualitative and partial,
they will at least give good guidance in more thorough future study.

\printbibliography

\end{document}